\documentclass[epj]{svjour}
\usepackage{amsmath,amssymb,graphicx}

\newlength{\colw}
\newcommand{\bra}{\langle}
\newcommand{\ket}{\rangle}
\newcommand{\braket}[1]{\langle#1\rangle}
\newcommand{\tr}{\operatorname{Tr}}

\newcommand{\One}{1\kern-4.5pt1}

\newcommand{\muhat}{\hat{\mu}}

\newcommand{\qq}{\bra qq\ket}

\newcommand{\cdeconf}{\cite{Hands:2006ve}}
\newcommand{\cquarkyonic}{\cite{Hands:2010gd}}
\newcommand{\cprev}{\cite{Hands:2006ve,Hands:2010gd,Cotter:2012mb}}
\newcommand{\ccurr}{\cite{Hands:2010gd,Cotter:2012mb}}
\newcommand{\cphases}{\cite{Cotter:2012mb}}

\begin{document}
\title{Phase transitions and gluodynamics in 2-colour matter at high density}
\titlerunning{Phase transitions in 2-colour matter}

\author{Tamer Boz\inst{1} \and Seamus Cotter\inst{1} \and Leonard
  Fister\inst{1} \and Dhagash Mehta\inst{2} \and Jon-Ivar Skullerud\inst{1}
}                     
%
%
\institute{Department of Mathematical Physics, National University of
  Ireland Maynooth, Maynooth, County Kildare, Ireland \and
Department of Physics, Syracuse University,
Syracuse, NY 13244, USA
}

\date{Received: date / Revised version: date}
%
\abstract{
We investigate 2-colour QCD with 2 flavours of Wilson fermion at
nonzero temperature $T$ and quark chemical potential $\mu$, with a
pion mass of 700 MeV ($m_\pi/m_\rho=0.8$).
From temperature scans at fixed $\mu$ we find that the critical
temperature for the superfluid to normal transition depends only very
weakly on $\mu$ above the onset chemical potential, while the
deconfinement crossover temperature is clearly decreasing with $\mu$.
We find indications of a region of superfluid but deconfined matter at
high $\mu$ and intermediate $T$.  The static quark potential
determined from the Wilson loop is found to exhibit a `string tension'
that increases at large $\mu$ in the `deconfined' region.  The
electric (longitudinal) gluon propagator in Landau gauge becomes
strongly screened with increasing temperature and chemical potential.
The magnetic (transverse) gluon shows little sensitivity to
temperature, and exhibits a mild enhancement at intermediate $\mu$
before becoming suppressed at large $\mu$.
} 
\maketitle
\section{Introduction}
\label{intro}

The properties and phase diagram of quantum chromodynamics (QCD) at
large baryon density remain largely unknown, despite substantial
theoretical efforts.  One important reason for this is the failure of
traditional Monte Carlo methods for lattice QCD at nonzero density,
due to the infamous sign problem.  While progress has been made in the
region of high temperature $T$ and moderate baryon chemical potential
$\mu_B$, the region of low $T$ and large $\mu_B$ remains inaccessible
to Monte Carlo simulations.  \footnote{Some progress has recently been made
using complex Langevin \cite{Aarts:2011ax,Aarts:2013bla} and other methods
\cite{Chandrasekharan:2012fk,Cristoforetti:2012su}, but neither have as yet been
shown to work for QCD.}

There are, however, QCD-like theories which are not affected by the
sign problem, at least for an even number of flavours $N_f$, among
them QCD with gauge groups SU(2) (QC$_2$D) \cprev,
or G$_2$ \cite{Maas:2012wr}, with nonzero isospin density
\cite{Kogut:2002zg} or with adjoint fermions \cite{Hands:2000ei,Hands:2001ee}.
These theories allow first-principles lattice simulations and may
hence be used as benchmarks for other methods that are not encumbered
by the sign problem, but which may involve uncontrolled
approximations.  Such methods may include model studies such as NJL
and quark--meson models (possibly augmented by a Polyakov loop
potential); effective theories valid for example for heavy quarks or
at high density; or functional methods such as the functional
renormalisation group (FRG), Dyson--Schwinger equations (DSEs) or
n-particle irreducible (nPI) methods, which rely on assumptions about
higher-order vertices.

Among these QCD-like theories, ${\rm QC_2D}$ is the simplest, both
mathematically and computationally; however, it shares important
properties with full QCD, in particular a hadronic phase and
deconfinement. Therefore, it has been adressed in various approaches
like analytic continuation methods at high temperature and low density
\cite{Giudice:2004se,Cea:2006yd,Cea:2009ba}, chiral effective theories
\cite{Kogut:1999iv,Kogut:2000ek,Splittorff:2000mm,Lenaghan:2001sd,Splittorff:2001fy,Kogut:2001if,Kogut:2002cm,Splittorff:2002xn,Dunne:2002vb,Dunne:2003ji,Brauner:2006dv,Kanazawa:2009ks,Kanazawa:2009en,Kanazawa:2011tt},
(Polyakov--)Nambu--Jona-Lasinio or (Polyakov--)Quark-Meson(-Diquark)
models
\cite{Giannini:1991ew,Kondratyuk:1992he,Rapp:1997zu,Ratti:2004ra,Sun:2007fc,Brauner:2009gu,Andersen:2010vu,Harada:2010vy,Zhang:2010kn,He:2010nb,Strodthoff:2011tz},
 Dyson--Schwinger equations \cite{Bloch:1999vk} and
lattice gauge theory \cite{Nakamura:1984uz,Hands:1999md,Kogut:2001na,Muroya:2002jj,Chandrasekharan:2006tz,Alles:2006ea,Lombardo:2008vc,Hands:2006ve,Hands:2010gd,Hands:2011ye}. From these studies a conjectured phase diagram in the
$(\mu,T)$ plane has emerged. At small values for chemical potential
and temperature, quarks are confined but their chiral symmetry is
broken which leads to a non-vanishing chiral condensate as in full
QCD, hence a hadronic phase. However, the Pauli--G\"{u}rsey symmetry
of ${\rm QC_2D}$ leads to a characteristic difference. In the chiral
limit for two flavours, the spontaneous symmetry breaking by the
chiral condensate necessitates 5 (massless) Nambu--Goldstone bosons,
three of which are the pions, while the remaining two are the diquark and
antidiquark. This pattern is realised approximately also for small
quark masses. The (anti-)diquarks play the role of the baryons, hence
having Bose--Einstein statistics in  ${\rm QC_2D}$. For vanishing
temperature but increasing chemical potential the Silver Blaze
property \cite{Cohen:2003kd} dictates that observables must not change
up to the threshold of the quark chemical potential $\mu=m_{\rm
  baryon}/N_c$, hence half of the pion mass for two colour
matter. Above this threshold, the different statistics of the baryons
lead to a qualitatively different phase structure compared to full
QCD: For $\mu> m_{\pi}/2$ the baryons condense and form a BEC of
diquarks $\langle qq\rangle \neq 0$, which possibly turns into a BCS condensate
at high densities, in contrast to the colour-superconductor picture in
full QCD. For increasing temperatures also ${\rm QC_2D}$ is expected
to exhibit a crossover transition to a deconfined phase, like full
QCD. This transition is signalled by the rise of the Polyakov-loop
$\langle L\rangle $. The structure of at least three distinct phases
is characterised by vanishing or nonvanishing diquark condensate
$\langle qq\rangle $ and Polyakov loop $\langle L\rangle $ and can be
summarised by 
\begin{itemize}
\item a vacuum or hadronic phase at low $T$ and $\mu$, characterised by
vanishing quark number density $n_q$ and with $\qq=0,
\braket{L}\approx0$;
\item a superfluid, confined (quarkyonic) region at
low $T$ and intermediate to large $\mu$, with
$\braket{L}\approx0,\qq\neq0$; 
\item a deconfined quark--gluon plasma at
high $T$, with $\braket{L}\neq0,\qq=0$; and possibly 
\item a deconfined,
superfluid region at large $\mu$ and intermediate $T$, with
$\braket{L}\neq0,\qq\neq0$.
\end{itemize}
A quarkyonic phase was first conjectured in \cite{McLerran:2007qj} in
the context of large $N_c$ and defined as confined and chirally
symmetric.  It was later found \cite{Kojo:2009ha} that chiral symmetry
may be broken in unconventional ways, and it seems more appropriate
instead to define quarkyonic as a state of matter where (weakly
interacting) quarks form the bulk degrees of freedom, but which
remains confined, ie all excitations are hadronic.  Quarkyonic matter
in QC$_2$D was studied in \cite{Brauner:2009gu}, and in
\cquarkyonic\ evidence for a
quarkyonic region was presented from lattice simulations.

In the present paper we extend the analysis of \cite{Cotter:2012mb},
where rough estimates for the phase boundaries between these regions
of ${\rm QC_2D}$ were presented. We study the phase transitions in
more detail and attempt to pinpoint their location. Since we are using
Wilson fermions and quite heavy quarks, we are not in a position to
study chiral symmetry directly.  An exploratory attempt to adress this
issue was made in \cphases.

We will also attempt to cast further light on the confining properties
of the theory at low temperature and the nature of the putative
deconfinement transition at high density, by computing the static
quark potential in the low-temperature region.


In contrast to quantities which may not be directly comparable between
theories, the effects of the medium on low order Green functions in 
QC$_2$D may provide a reliable guideline to full QCD. Quark and
gluon correlation
functions are of great interest, as the theory can be fully expressed
in terms of these. Propagators play a predominant role, in particular
in continuum descriptions, and in some cases their behaviour suffices
to shed light on the critical physics of the phase diagram, e.g.\ the
deconfinement transition
\cite{Braun:2007bx,Marhauser:2008fz,Braun:2010cy,Fister:2013bh}.  In
this paper, we will study how the gluon propagator responds to both
temperature and quark chemical potential.

In Sec.~\ref{sec:simulation} we set out the details of our lattice
simulations, including the action, parameters and lattice volumes
used.  Then, in Sec.~\ref{sec:transitions} we study the superfluid to
normal and deconfinement transition by performing a temperature scan
at 3 different values of the chemical potential.  The response of the
static quark potential to $\mu$ is investigated in
Sec.~\ref{sec:potential}, while in Sec.~\ref{sec:gluon} results for
the gluon propagator are reported.  Appendix~\ref{appendix} contains
some further details about the diquark source extrapolation of the
superfluid condensate.  Preliminary results for the gluon
propagator have been reported in
\cite{Skullerud:2008wu,Skullerud:2009hq}, and for the static quark
potential in \cite{Cotter:2012ny}.

\section{Simulation details}
\label{sec:simulation}

We use a standard Wilson gauge action with two flavours of unimproved
Wilson fermion, with the addition of a diquark source term to lift the
low-lying eigenmodes and allow a controlled study of diquark
condensation effects.  Further details about the action and the
simulation method can be found in \cprev.  The results obtained will
depend on the diquark source $j$; in the end the $j\to0$ limit must be
taken to obtain `physical' results.\footnote{In cases where model
  studies could be carried out with $j\neq0$ one might also compare
  directly results for nonzero $j$; however, most other studies will
  not contain any explicit diquark source term, so the $j\to0$ limit
  is crucial.}

\begin{table*}
\begin{tabular}{ll|*{15}r}
\hline
&& \multicolumn{15}{c}{$N_\tau$} \\ \cline{3-17}
$\mu a$ & $ja$ & 20 & 18 & 16 & 15 & 14 & 13 & 12 & 11
 & 10 & 9 & 8 & 7 & 6 & 5 & 4\\ \hline
0.35 & 0.02 & & & 270 & & & 370 & 250 & 270 & 510 & 560 & 800 & 520 & 1100 & 300 & 250 \\
0.35 & 0.04 & & & 250 & & 270 & & 4550 & 710 & 500 & 500 & 675 & 510 & 900 & 300 & 250
 \\ \hline
0.4 & 0.02 & & & 270 & & & 250 & 500 & 500 & 500 & 550 & 1000 & 250 &
 250 & 300 \\
0.4 & 0.03 & & & & & & & & 250 & 280 & 300 \\
0.4 & 0.04 & & & & & & 1020 & 1080 & 1050 & 1000 & 1000 & 1050
 & 1000 & 1000 & 1000 & 1000 \\ 
0.4 & 0.05 & & & & & & & & 250 & 240 & 216 \\ \hline
0.5 & 0.02 & & & 280 & & & & 512 & 250 & 255 & 275 & 1000 & 250 & 300 \\
0.5 & 0.03 & & & & & & & & 280 & 270 & 280 & 1000 \\
0.5 & 0.04 & 3075 & 3020 & 2570 & 3835 & 3340 & 3200 & 3240 & 2620 
 & 1000 & 1000 & 1050 & 1200 & 1000 & 1200 & 1000 \\
0.5 & 0.05 & & & & & & & & 300 & 270 & 270 \\ \hline
0.6 & 0.02 & & & 350 & & 320 & & 310 & 300 & 300 & 300 & 1000
 & 300 & 300 \\
0.6 & 0.03 & & & & & & & & 290 & 280 & 280 \\
0.6 & 0.04 & 4200 & & 2300 & 2210 & 1840 & 1000 & 1128 & 1005 & 1000 
 & 1040 & 1200 & 1020 & 1000 \\
0.6 & 0.05 & & & & & & & & 255 & 300 & 300 \\
\hline
\end{tabular}
\caption{Number of trajectories for different temperatures,
  $T=1/(aN_\tau)$, chemical potentials $\mu$ and diquark sources $j$.
  All trajectories have average length 0.5, and the spatial lattice
  size is $N_s=16$ in all cases.}
\label{tab:Tscans}
\end{table*}

We use the same parameters as in \ccurr, namely
$\beta=1.9,\kappa=0.168$, corresponding to a lattice spacing
$a=0.178(6)$fm and a pion mass $am_\pi=0.645(8)$, or $m_\pi=717(25)$MeV.  The lightest
baryon, the scalar diquark, is degenerate with the pion in the vacuum,
and at zero temperature we therefore expect an onset transition to a
superfluid phase at $m=m_\pi/2$.  This has been corroborated in
previous simulations \cprev.

In addition to the ensembles used and described in \cphases, we have
generated gauge configurations on $16^3\times N_\tau$ lattices with
$N_\tau=$4--20, in order to study in detail the thermal transitions at
$a\mu=0.35, 0.4, 0.5$ and 0.6.  The details of these ensembles are
given in table~\ref{tab:Tscans}.  For most temperatures, two diquark
sources $ja=0.02$ and 0.04 have been used, enabling us to perform a
linear extrapolation to the $j=0$ limit.  In the region of the
superfluid to normal transition, where a linear extrapolation is known
to be invalid, two additional $j$-values have been added to allow for
a controlled extrapolation.

\section{Phase transitions}
\label{sec:transitions}

\subsection{Superfluid to normal transition}
\label{sec:superfluid}

\begin{figure}[tb]
\includegraphics*[width=\colw]{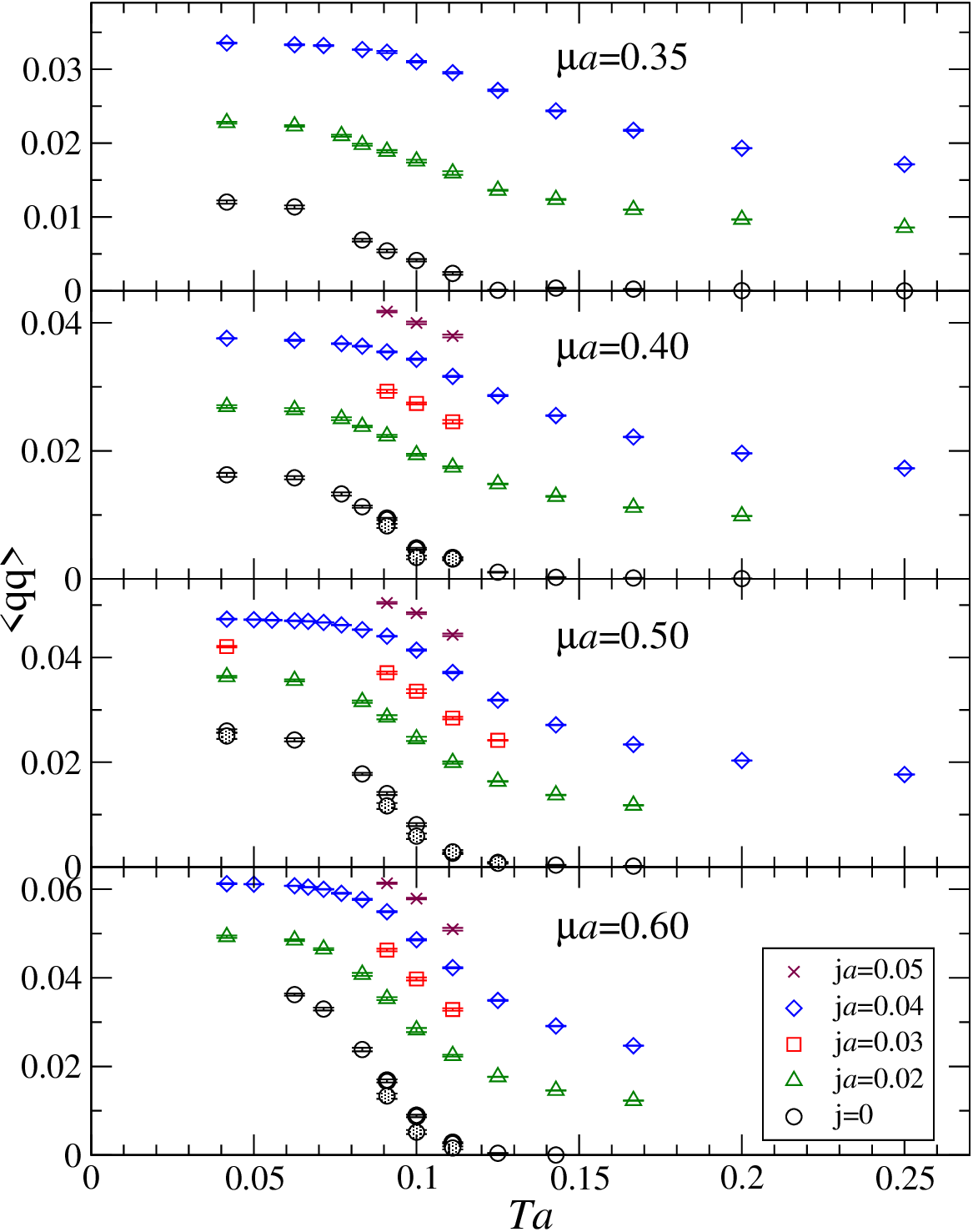}
\caption{Diquark condensate $\qq$ as a function of temperature $T$ for
  chemical potential $\mu a=0.35, 0.4, 0.5, 0.6$ (top to bottom).  The
circles are data extrapolated to $j=0$ using a linear Ansatz for
$ja\leq0.04$; the shaded circles denote the results of a linear
extrapolation using $j=0.02,0.03$ only.}
\label{fig:diquark}
\end{figure}

Figure~\ref{fig:diquark} shows the order parameter for superfluidity,
the (unrenormalised) diquark condensate
\begin{equation}
\qq = \bra\psi^{2tr}C\gamma_5\tau_2\psi^1
  -\bar\psi^1C\gamma_5\tau_2\bar\psi^{2tr}\ket\,,
\label{def:qq}
\end{equation}
as a function of the temperature $T$, for $\mu a=0.35, 0.4,0.5$ and 0.6.
Also shown are the results of a linear extrapolation to $j=0$.  Note
that renormalisation will only amount to multiplying these data by a
$\mu$- and $T$-independent constant, and will hence not change the $\mu$-
and $T$-dependence of the results.

We can
clearly observe a transition from a superfluid phase, characterised by
$\qq\neq0$, at low temperature, to a normal phase with $\qq=0$ at high
temperature, with a transition in the region $0.08\lesssim
Ta\lesssim0.12$ for all three values of $\mu$.  In an attempt to more
precisely locate
the transition, we have performed simulations with 4 different
$j$-values in the transition region, which may allow for a controlled
$j\to0$ extrapolation. However, none of the functional forms we have
used give satisfactory results. Details of these extrapolations are given in
Appendix~\ref{appendix}.

With our current data at only a single volume we are not in a position
to determine the order of the transition, although it would be
expected to be a second order phase transition in the universality
class of the 3-dimensional XY model. This could be tested by
attempting a universal fit of the data at $j\neq0$ in the transition
region to a scaling form given by the appropriate critical exponents:
note that at fixed $\mu$, the diquark source $j$ would be the magnetic
field variable in this scaling function.  This would also provide an
alternative method for determining $T_s$.

Because of the uncertainties in the $j\to0$ extrapolation in the
critical region, we have estimated the critical temperatures $T_s$
for the superfluid to normal transition by determining the inflection
points for $\qq$ at $ja=0.02$ and 0.04, and extrapolated the resulting
values to $j=0$ using a linear Ansatz.  The results are given in
table~\ref{tab:Ts}.
\begin{table}
\begin{center}
  \begin{tabular}{c|rrrr}
    $a\mu$ & 0.35 & 0.40 & 0.50 & 0.60 \\\hline
$aT_s(0.04)$ & 0.121(6) & 0.108(2) & 0.111(5) & 0.102(6) \\
$aT_s(0.02)$ & 0.097(16) & 0.096(5) & 0.097(2) & 0.093(5) \\ \hline
$aT_s$ & 0.073(24) & 0.084(8) & 0.083(5) & 0.083(6) \\
$T_s$ (MeV) & 82(27) & 94(9) & 93(6) & 93(7)
  \end{tabular}
  \caption{Inflection points $T_s(j)$ for $\qq(T)$ at $ja=0.04,0.02$
    and critical temperature $T_s$ obtained from extrapolating
    $T_s(j)$ to $j=0$.  The uncertainties are estimates of the
    systematic uncertainty in determining the inflection points and in
    the $j\to0$ extrapolation.}
  \label{tab:Ts}
\end{center}
\end{table}
We see that $T_s$ is remarkably constant over the whole range of
$\mu$-values considered.  The indications are that the transition
happens at a somewhat lower temperature at $\mu a=0.35$, but this
point is already very close to the onset from vacuum to superfluid at
$T=0$, $\mu_oa=m_\pi a/2=0.32$, suggesting that $T_s(\mu)$ rises very
rapidly from zero at $\mu=\mu_o$ before suddenly flattening off.

\subsection{Deconfinement transition}
\label{sec:deconfine}

The Polyakov loop $\braket{L}$ serves as the traditional order
parameter for deconfinement in gauge theories, with $\braket{L}\neq0$
signalling the transition to a deconfined phase.  Strictly speaking,
$\braket{L}$ is never zero in a theory with dynamical fermions, but it
typically increases with temperature from a very small value in a
fairly narrow region, which may be identified with the deconfinement
transition region.

We will here take the pragmatic view in which QC$_2$D, like QCD, is
considered to be confining at low $T$ and $\mu$.  This is also
reflected in the behaviour of the static quark potential, which will
be studied in the following section: it rises linearly at intermediate
distances, before string breaking sets in.

\begin{figure*}
\includegraphics*[width=\textwidth]{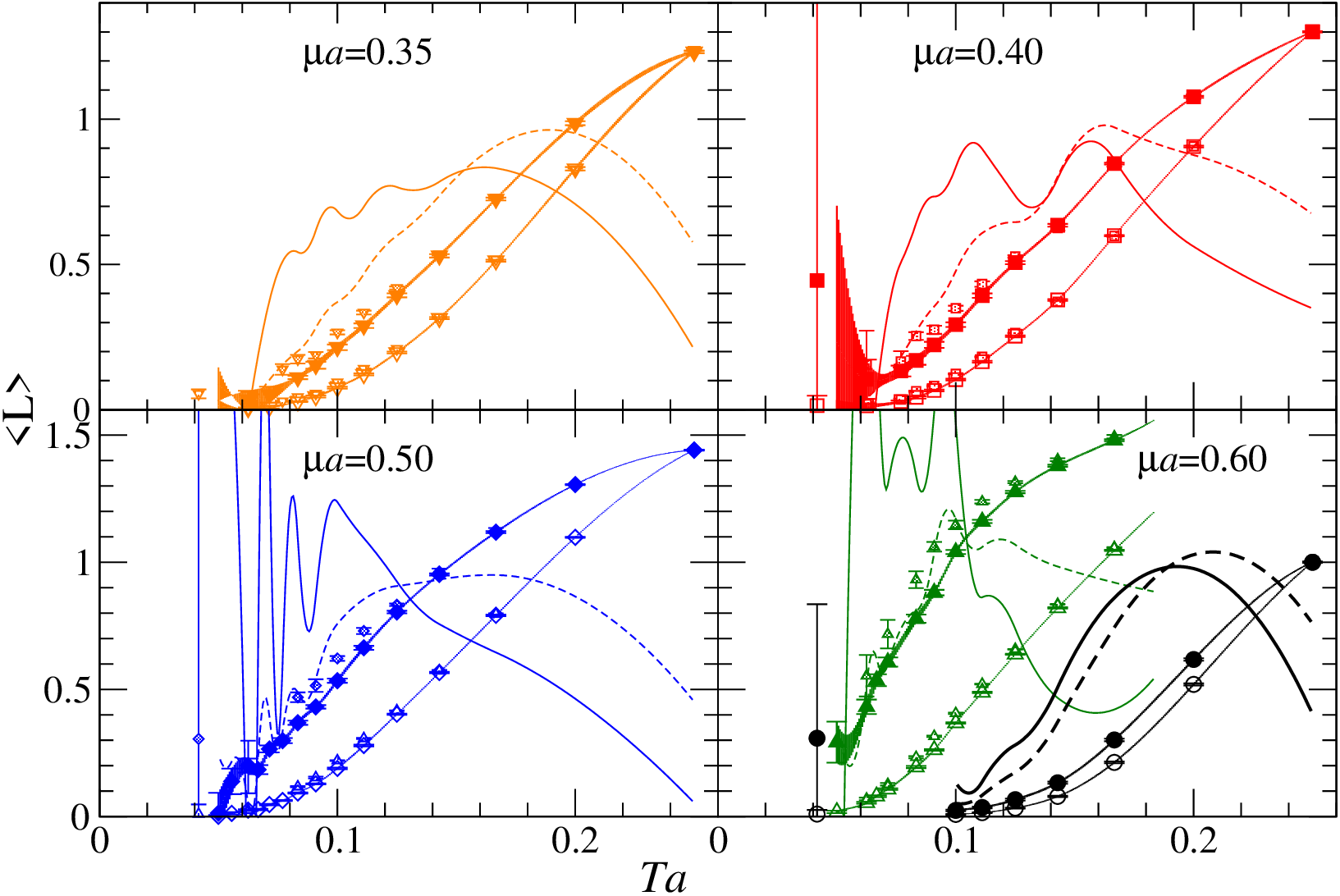}
\caption{The renormalised Polyakov loop $\braket{L}$ as a function of
  temperature $T$ for $ja=0.04$ and $\mu a=0.35,0.4,0.5, 0.6$, with
  two different renormalisation schemes: Scheme A (solid symbols) and
  Scheme B (open symbols), see text for details.  The solid (dashed)
  lines are the derivatives of cubic spline interpolations of the data
  points for Scheme A (B).  The smaller, shaded symbols are results
  for $ja=0.02$.  The black circles and thick lines in the bottom
  right panel are the $\mu=j=0$ results from \cphases.}
\label{fig:polyakov-allmu}
\end{figure*}
Unlike the diquark condensate, the renormalisation of the Polyakov
loop does depend on temperature; specifically, the relation between
the bare Polyakov loop $L_0$ and the renormalised Polyakov loop $L_R$
is given by
\begin{equation}
L_R(T,\mu)= Z_L^{N_\tau}L_0(\frac{1}{aN_\tau},\mu)\,.
\label{eq:polyakov-renorm}
\end{equation}
In order to investigate the sensitivity of our results to the
renormalisation scheme, we have used two different conditions to
determine the constant $Z_L$,
\begin{align*}
\textbf{Scheme A} & \qquad & L_R(T=\frac{1}{4a},\mu=0) &= 1\,,\\
\textbf{Scheme B} & \qquad & L_R(T=\frac{1}{4a},\mu=0) &= 0.5\,.
\end{align*}
Scheme A is the scheme that was already used in \cphases.
Figure~\ref{fig:polyakov-allmu} shows $\braket{L}$ evaluated in both
schemes, as a function of temperature.  The Scheme B data have been
multiplied by 2 to ease the comparison with the Scheme A data.  Also
shown are cubic spline interpolations of the data and the derivative
of these interpolations, with solid lines corresponding to Scheme A
and dotted lines to Scheme B.
 
At all $\mu$, we see a transition from a low-temperature confined
region to a high-temperature deconfined region.  In contrast to the
diquark condensate, we see a clear, systematic shift in the transition
region towards lower temperatures as the chemical potential increases.
For all four $\mu$-values, the Polyakov loop shows a nearly linear
rise as a function of temperature in a broad region, suggesting that 
the transition is a smooth crossover rather than a true phase
transition.  This is reinforced by the difference between Scheme A and
Scheme B, with the crossover occuring at higher temperatures in Scheme
B.  At $\mu=0$, the difference between the two schemes is small, but
increases with increasing $\mu$, suggesting a broadening of the
crossover.

Because of the smaller value of $Z_L$, our results for Scheme B are
considerably less noisy than those for Scheme A.  For this reason, we
choose to define the crossover region to be centred on the inflection
point from Scheme B, with a width chosen such that it also encompasses
the onset of the linear region from Scheme A. 
%
Our summary of transition temperatures taken from the $ja=0.04$ data
is given in table~\ref{tab:deconf}.  From
Fig.~\ref{fig:polyakov-allmu} we see that at low $T$, the value of $\braket{L}$
increases as $j$ is reduced, and at $\mu a=0.6$, the
crossover region will most likely move to smaller $T$ in the $j\to0$
limit.  However, we do not have sufficient statistics for $ja=0.02$ at
low $T$ to make any quantitative statement about this.
\begin{table}
\begin{center}
\begin{tabular}{l|cc}
$\mu a$ & $T_da$ & $T_d$ (MeV) \\ \hline
0.0 & 0.193(20) & 217(23) \\
0.35 & 0.140--0.220 & 157--247 \\
0.40 & 0.108--0.200 & 121--225 \\
0.50 & 0.080--0.200 & 90--225 \\
0.60 & 0.060--0.135 & 67--152
\end{tabular}
\end{center}
\caption{Estimates for the deconfinement crossover temperature $T_d$
  from the Polyakov loop at $ja=0.04$.  The $\mu=0$ result is taken
  from \cphases.}
\label{tab:deconf}
\end{table}

\section{Static quark potential}
\label{sec:potential}

The potential between two static quarks (or a quark--antiquark pair),
and in particular its asymptotic behaviour at large separations, has
traditionally been taken as the tell-tale indicator, or even
definition, of confinement of quarks \cite{Wilson:1974sk}.  A linearly
rising potential has been observed in numerous lattice simulations,
and has also formed the basis of successful phenomenological
descriptions of bound states of heavy quarks.  In QCD with dynamical
quarks, the string will break at a finite distance, but at
intermediate distances a linear rise can still be observed.

At high temperature, the potential is expected to exhibit Debye
screening, and this has been observed in numerous calculations of the
quark--antiquark free energy using Polyakov loop correlators.
However, it is not yet clear how this quantity relates to the
(complex) potential that appears in effective theories of heavy
quarkonia at high temperature \cite{Laine:2006ns,Brambilla:2008cx,Beraudo:2007ky,Beraudo:2010tw}.  Very recently, the
static quark potential has also been determined from Wilson loops at
high temperature \cite{Bazavov:2012bq}; this does not show any
screening for $T\lesssim T_c$.

There has also been some recent progress in determining the potential
between heavy (finite mass) quarks at zero \cite{Kawanai:2011jt} and non-zero
\cite{Allton:2012ki} temperature.  Some properties of bound states of
heavy quarks in QC$_2$D at nonzero temperature and density were
reported in \cite{Hands:2012yy}; a potential model description should
reproduce these results.  Here we compute the static quark potential
from Wilson loops for our lowest temperature, the $12^3\times24$ lattices.

\begin{figure}[tb]
\includegraphics*[width=\colw]{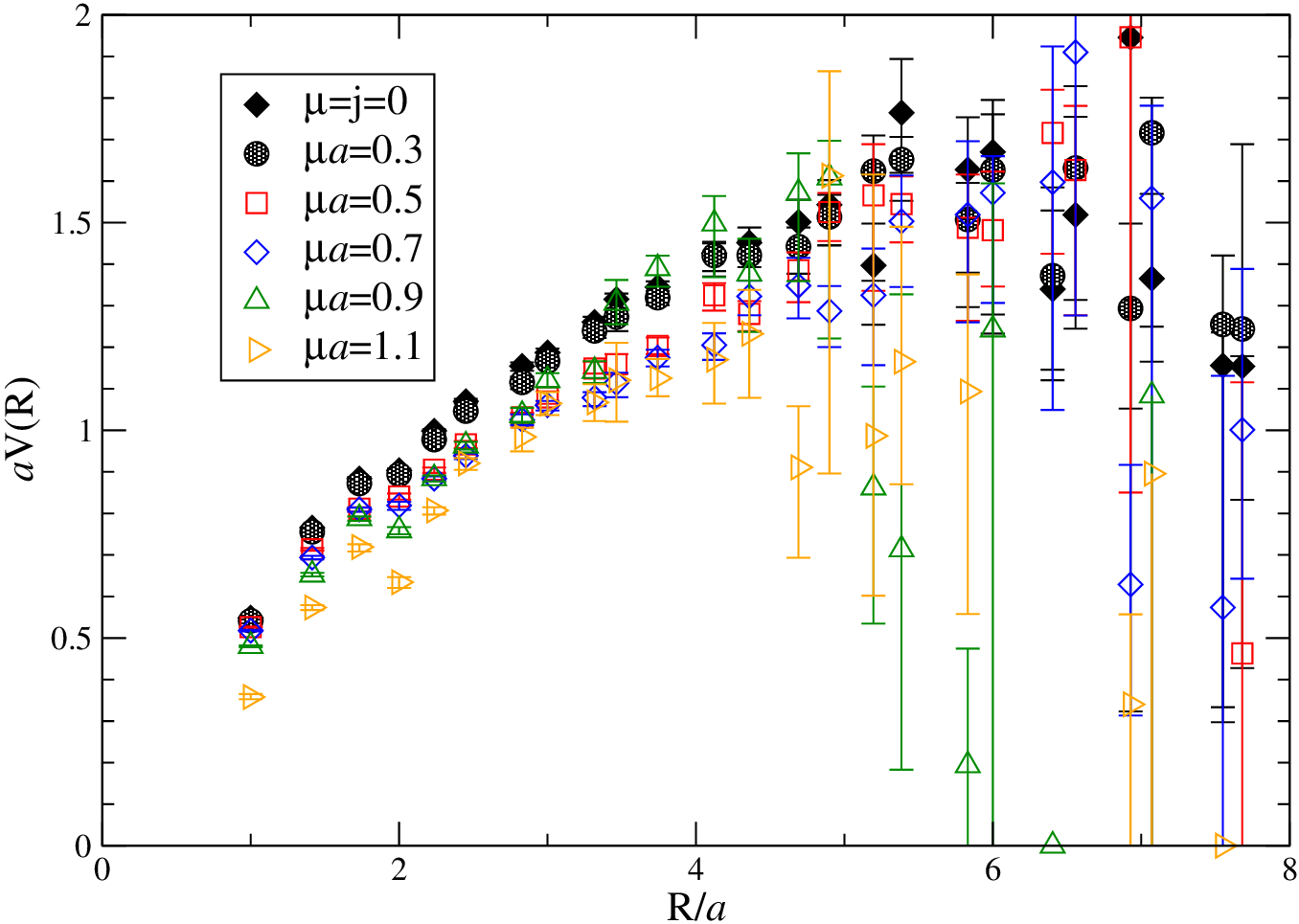}
\caption{The static quark potential computed from the
  Wilson loop, for the $12^3\times24$ lattice and different values of
  $\mu$, with $ja=0.04$.}
\label{fig:potential}
\end{figure}

In fig.~\ref{fig:potential} we show the static quark potential
computed from the Wilson loop at $N_\tau=24$, for $\mu a=0.3, 0.5,
0.7, 0.9$ and 1.1.  The data have been obtained from fits to the
asymptotic form of the Wilson loop, $W(r,\tau)=C\exp(-V(r)\tau)$,
assuming ground state dominance for $\tau\geq3$. We find that as we
enter the superfluid region, the 
potential becomes slightly flatter, but as $\mu$
is increased further no additional screening is observed, and at $\mu
a=0.9$, which according to our analysis of the Polyakov loop
should be in the deconfined region, the potential is consistent with
the $\mu=0$ potential.  This is in qualitative agreement with the pattern that
was already observed in \cdeconf.

\begin{figure}[tb]
\includegraphics*[width=\colw]{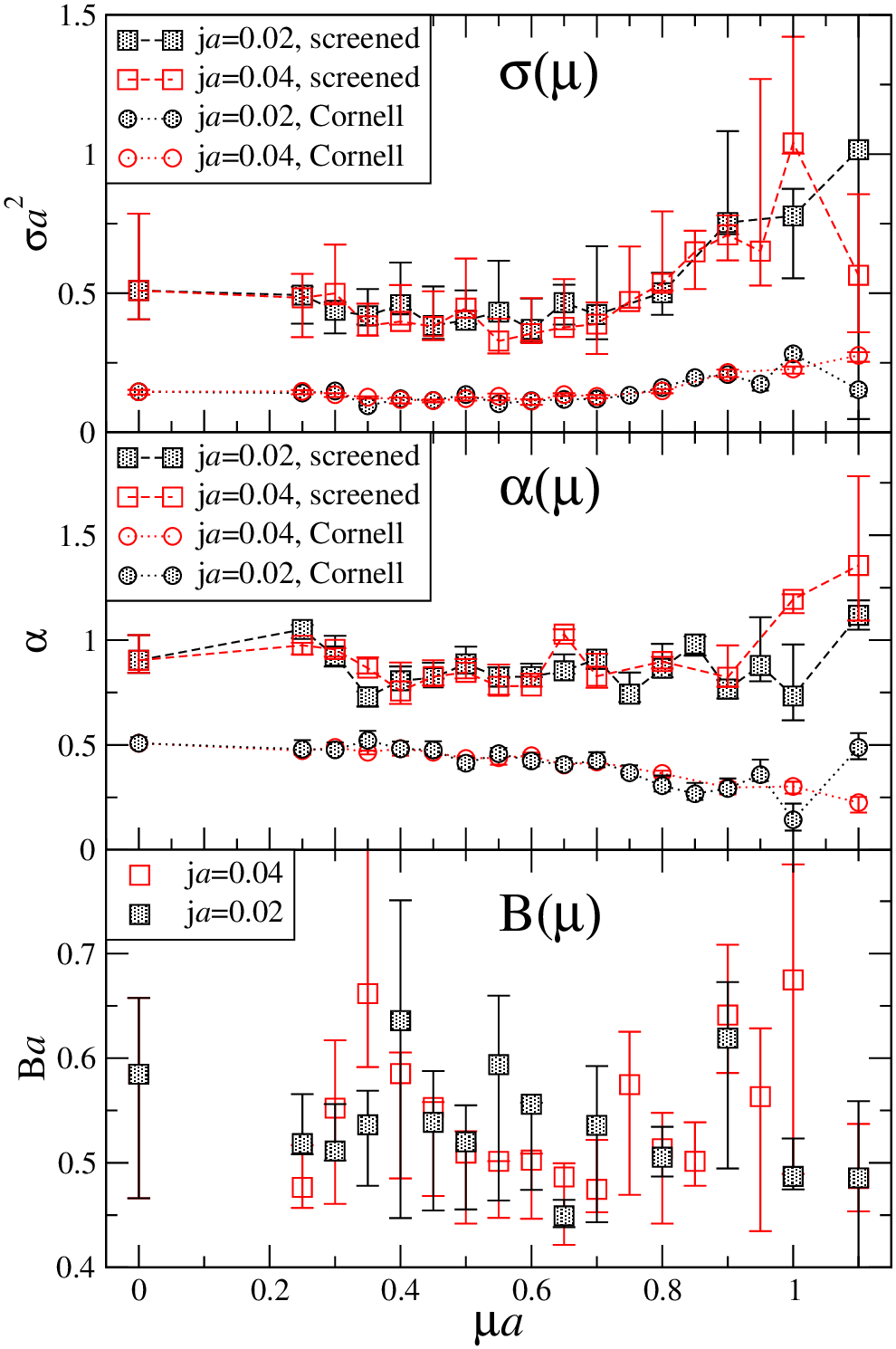}
\caption{The parameters $\sigma$ (top), $\alpha$ (middle) and $B$
  (bottom) in the screened
  Cornell potential \eqref{eq:screened_Cornell}, as a function of the chemical
  potential $\mu$, for the $12^3\times24$ lattice.  Also shown are the
  results for $\sigma$ and $\alpha$ using the pure Cornell potential \eqref{eq:Cornell}.}
\label{fig:string}
\end{figure}
To quantify the variation of the static quark potential with $\mu$, we
have performed a fit to the Cornell potential,
\begin{equation}
V(r) = C(\mu,j) + \sigma(\mu,j)r + \frac{\alpha(\mu,j)}{r}\,,
\label{eq:Cornell}
\end{equation}
and a screened potential with an exponential term to allow for screening effects and a decay of the linear part,
\begin{equation}
V(r) = C(\mu,j) + \frac{\sigma(\mu,j)r}{B(\mu,j)}e^{-Br} + \frac{\alpha(\mu,j)}{r}\,,
\label{eq:screened_Cornell}
\end{equation}
for each $\mu$ and $j$.  The results for the ``string tension''
$\sigma$ and the Coulomb factor $\alpha$ are shown in
fig.~\ref{fig:string}.  For $\mu a<0.6$ we do not see any systematic
trend, but for larger $\mu$ we find that $\sigma$ increases with
$\mu$.  We also see that there is no significant dependence on the
diquark source $j$. The exponential term appears to be insensitive to
$\mu$, and is clearly non-zero already at $\mu=0$, and should
therefore not be interpreted as a screening mass.  For the same
reason, the values for $\sigma$ and $\alpha$ from the two fits cannot
be directly compared.

In \cite{Bazavov:2013zha} it was observed that the static quark
potential extracted from Wilson loops in QCD at $\mu=0$ is more weakly
screened at intermediate temperature than the free energy determined
from Polyakov loop correlators.  Our results at large $\mu$ appear
consistent with this.  
There are, however, several possible complicating factors:
\begin{enumerate}
\item The transition may be to a medium characterised by long-range
  interactions, rather than by colour screening.  If that is the case,
  one would expect typical correlation lengths to also grow with
  $\mu$.  Reconciling an unscreened potential with a nonzero
  Polyakov loop remains a challenge, though.
\item The quark--antiquark potential may be screened at large distances,
  but this is not observed in the Wilson loop because of poor overlap
  with the relevant states.  This corresponds to the standard scenario
  at low temperature, where explicit mesonic states must be introduced
  to observe string breaking \cite{Bali:2005fu}.  Clearly, our
  observation of a linearly rising potential (area law for the Wilson
  loop) does not prove that the medium is confining.  The increasing
  slope could however be indicative of a large internal energy for
  static quark--antiquark pairs at intermediate distances.

In \cite{Hands:2012yy} it was found that the binding energy of heavy
quarkonia in QC$_2$D increases up to $\mu a\approx0.7$, and decreases
again beyond that.  This appears to run counter to the conjecture
above, in which the quarkonium might be expected to become more
strongly bound.
\item A more pessimistic scenario is that the whole region of $\mu
  a>0.6$ could be dominated by lattice artefacts.  Unfortunately our
  previous data at $\beta=1.7$ \cdeconf\ are probably outside the
  scaling region so a comparison with those is likely to be not very
  revealing.  Simulations at smaller $a$, which are underway, should
  confirm or rule out this scenario.
\end{enumerate}

Computing the static quark potential
using Polyakov loop correlators rather than Wilson loops might yield
further insight into this issue. However, the Polyakov loop
correlator suffers from the same signal to noise problem at low $T$
(large $N_\tau$) as the Polyakov loop itself, and we have not been
able to obtain any signal for $N_\tau=24$ except for $\mu
a\gtrsim0.9$.
Results for higher temperatures using both the Wilson loop and
Polyakov loop correlators will be presented in a future publication.

\section{Gluon propagator}
\label{sec:gluon}

\begin{figure}
\includegraphics*[width=\colw]{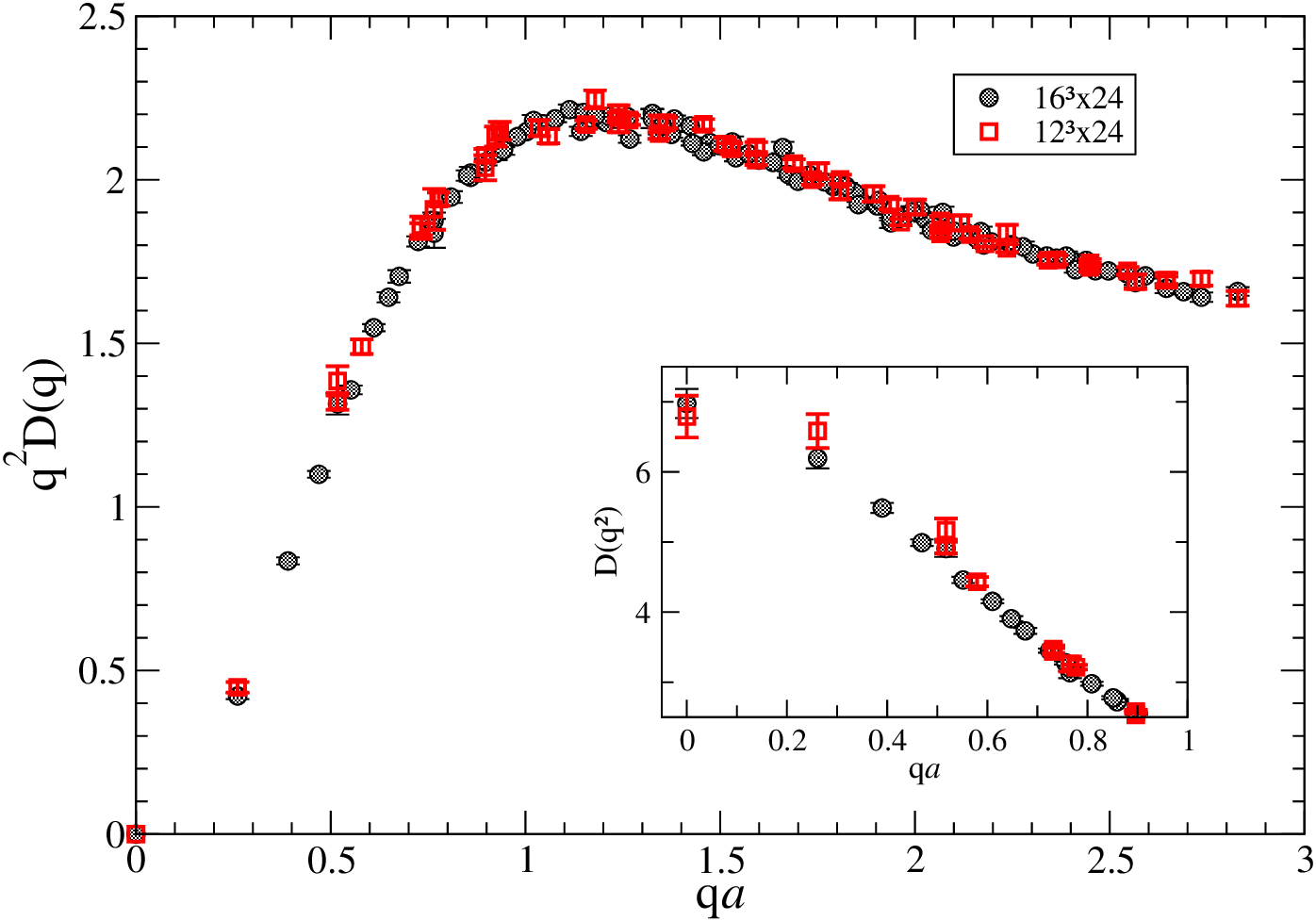}
\caption{The gluon dressing function in the vacuum from the
  $16^3\times24$ and $12^3\times24$ lattice.  The inset shows the
  gluon propagator for infrared momenta.  A cylinder cut has been
  applied to the data to reduce lattice spacing artefacts.}
\label{fig:gluon-vacuum}  
\end{figure}

\begin{figure*}
\includegraphics*[width=\colw]{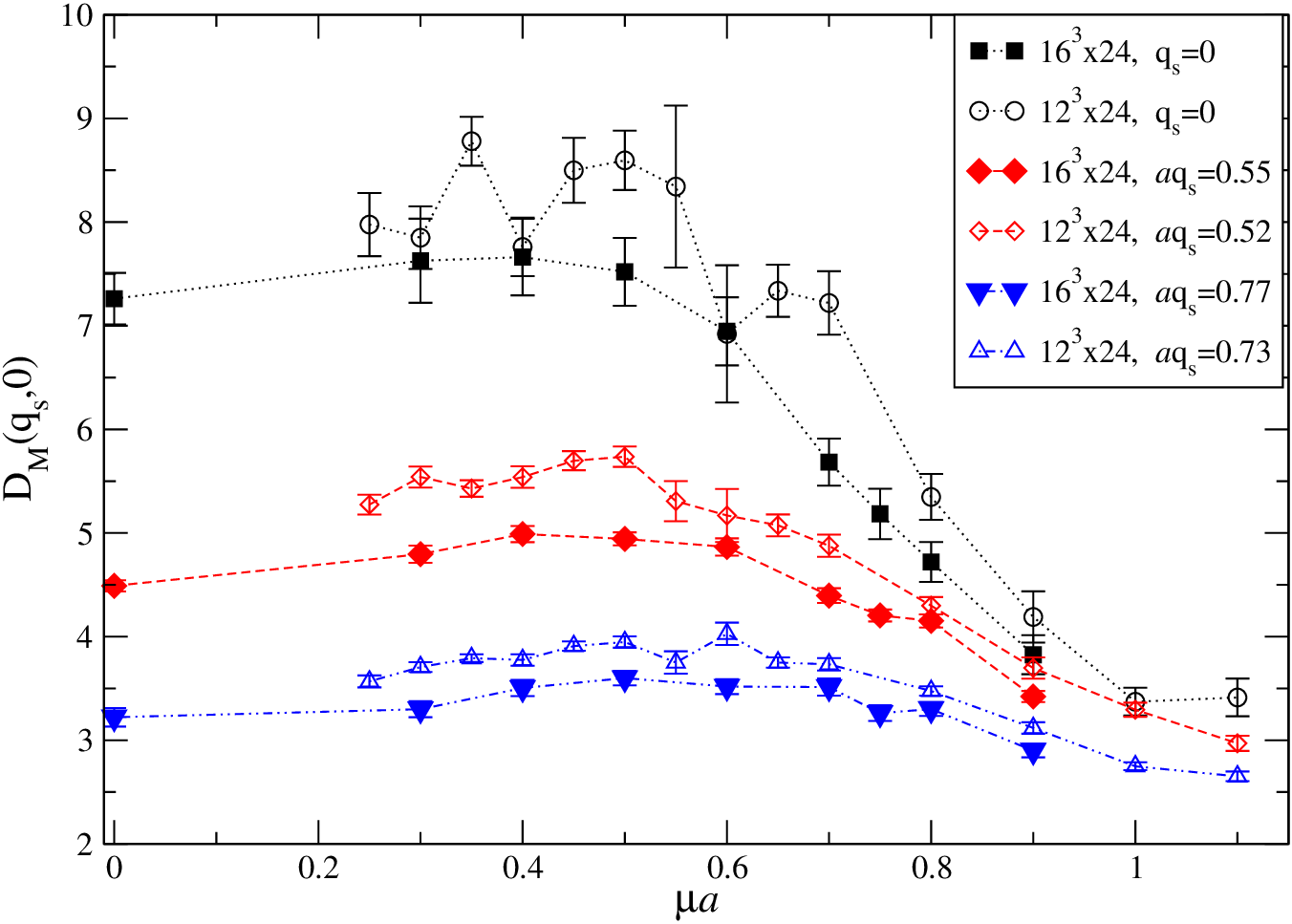}
\includegraphics*[width=\colw]{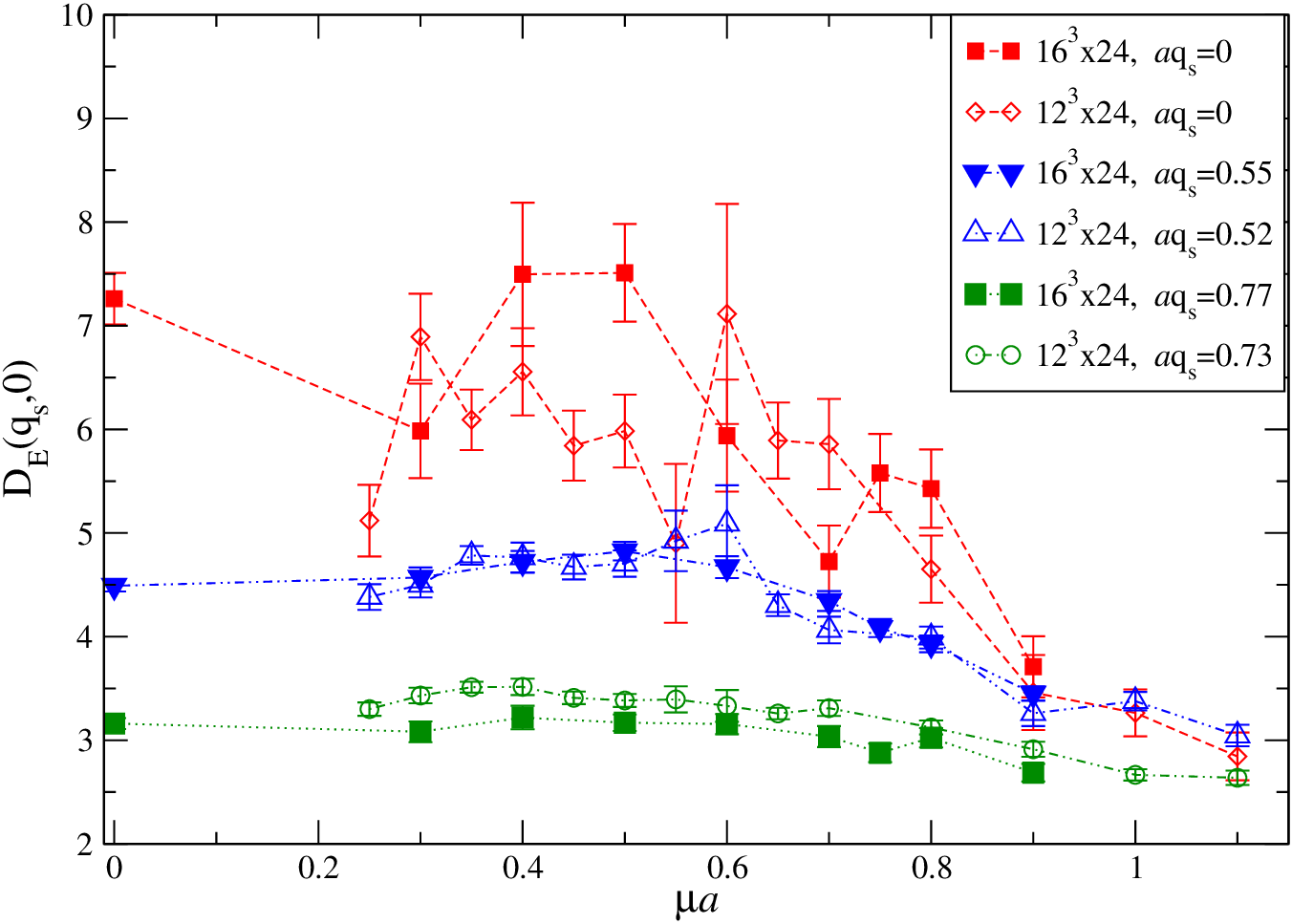}\\
\includegraphics*[width=\colw]{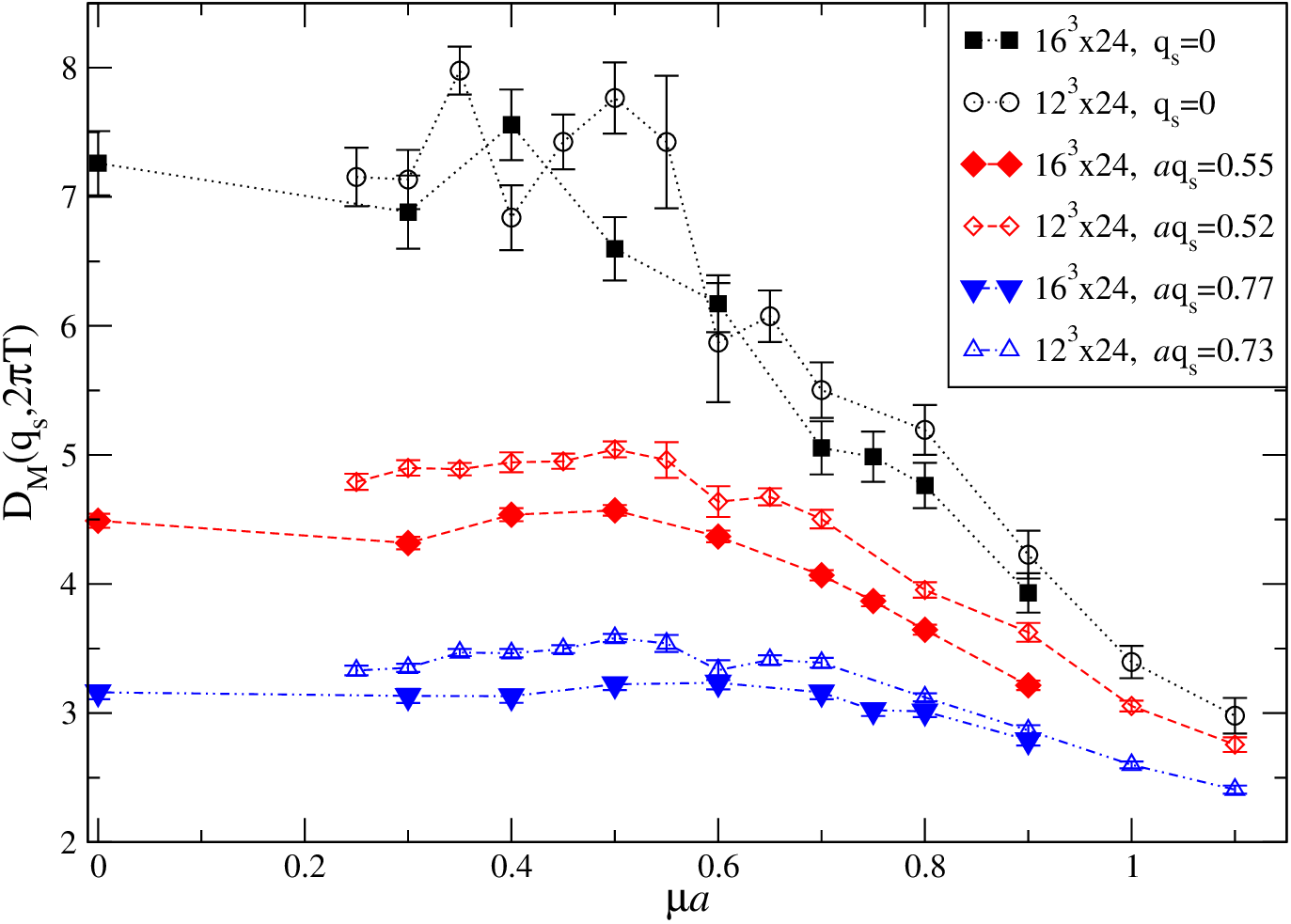}
\includegraphics*[width=\colw]{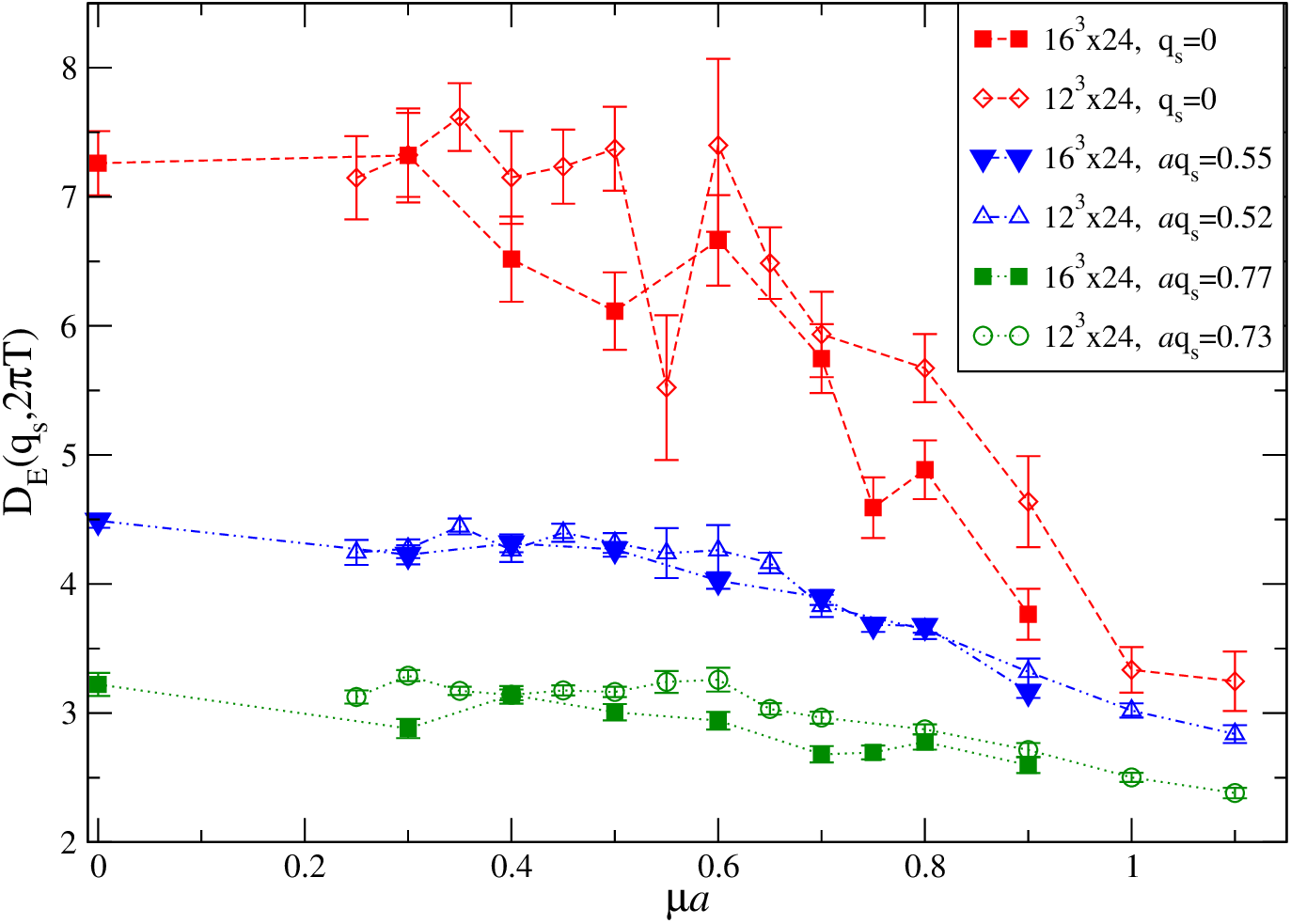}
  \caption{The zeroth (top) and first (bottom) Matsubara mode of the
    magnetic (left) and electric (right) gluon propagator as a
    function of chemical potential $\mu$ for selected values of the
    spatial momentum $q_s$, for $N_\tau=24$, different spatial volumes.}
\label{fig:gluon-compare}  
\end{figure*}

\begin{figure}
\includegraphics*[width=\colw]{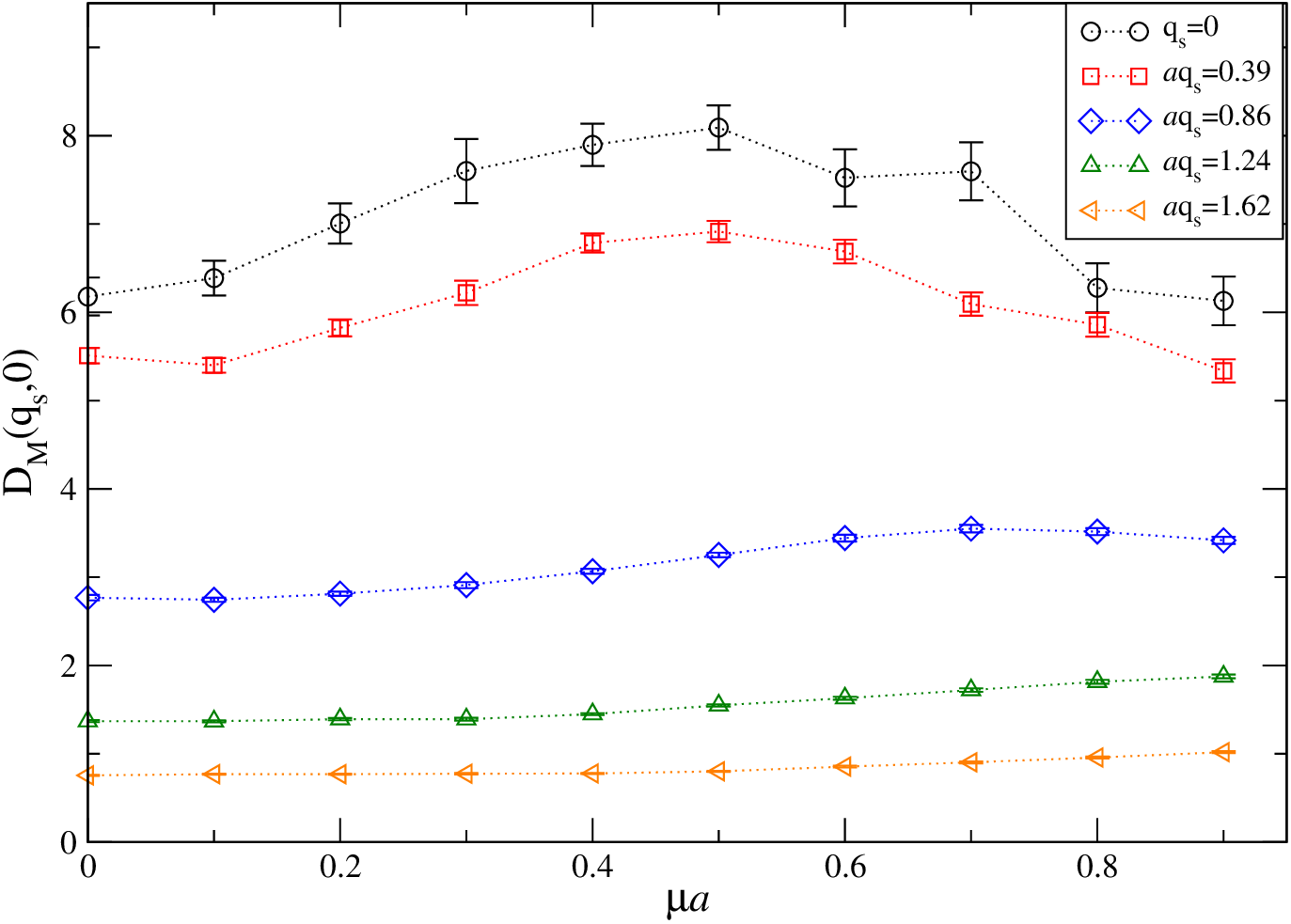}\\
\includegraphics*[width=\colw]{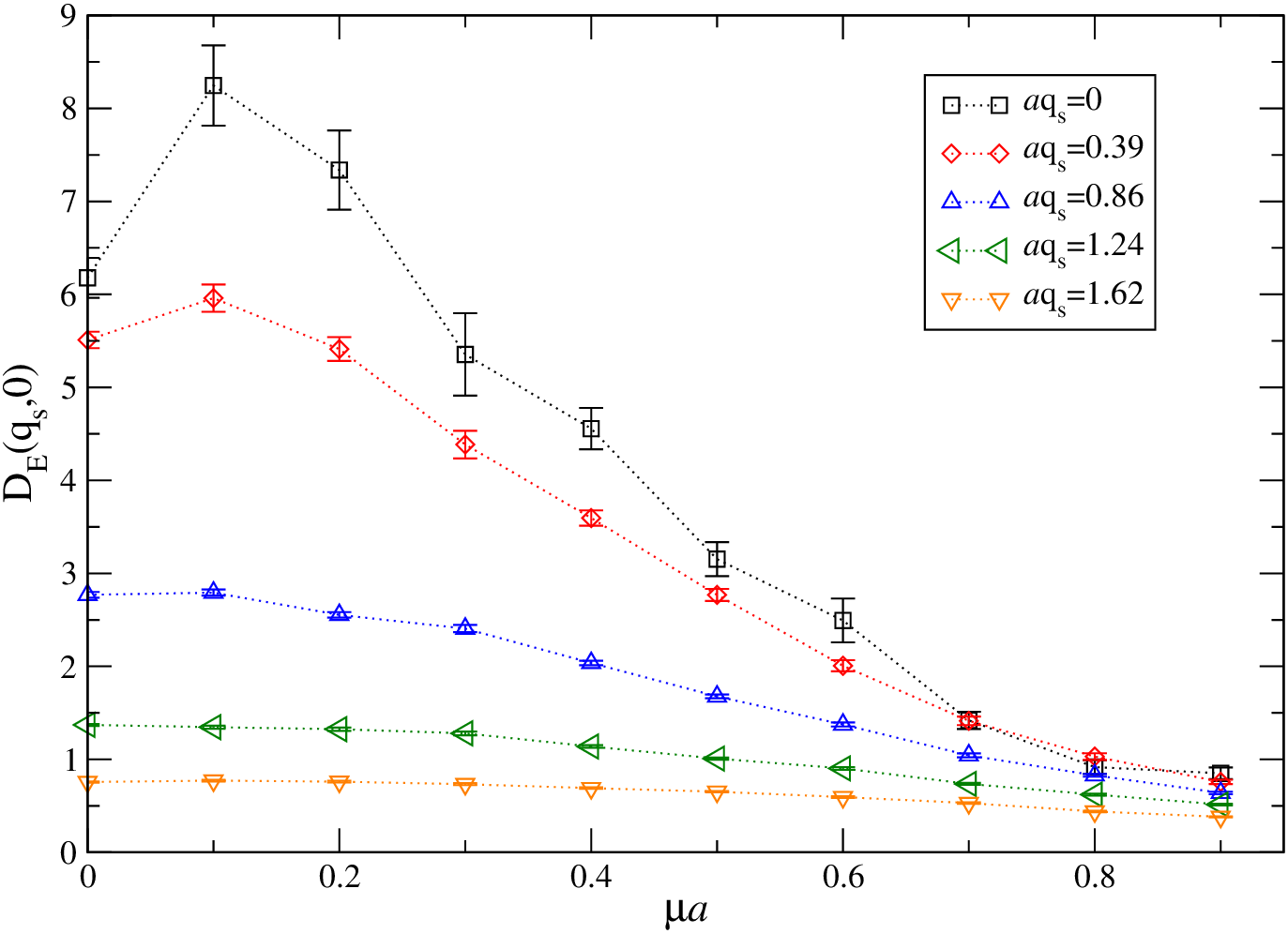}
  \caption{The zero Matsubara mode of the magnetic (top) and electric
    (bottom) gluon propagator as a
    function of chemical potential $\mu$ for selected values of the
    spatial momentum $q_s$, for the $16^3\times8$ lattice.}
\label{fig:gluon-16x8}
\end{figure}

In this section we extend previous studies \cite{Hands:2006ve,Skullerud:2008wu,Skullerud:2009hq} and present results for in-medium
gluon propagators, where we study the dependence on both parameters,
chemical potential and temperature. In Landau gauge only the
transverse part of the vacuum propagator is non-zero. However, the
external parameters break manifest Lorentz invariance, hence the gluon
propagator $D$ must be decomposed into chromoelectric and
chromomagnetic modes, $D_E$ and $D_M$, respectively,
\begin{equation}
D_{\mu\nu}\ =\ P_{\mu \nu}^{M} D_M \, + \, P_{\mu \nu}^{E} D_E
\,,
\label{eq:gluon_decomposition}
\end{equation}
where the individual dependence on (discrete) temporal and spatial
momenta has been omitted. The projectors on the longitudinal and
transversal spatial subspaces, $P_{\mu \nu}^{E}$ and $P_{\mu\nu}^{M}$,
are defined by 
\begin{align}
P_{\mu \nu}^{M} (q_0,\vec{q\,})
 &=  \left(1-\delta_{0\mu} \right)\left(1-\delta_{0\nu} \right)
    \left(\delta_{\mu\nu} -\frac{q_\mu q_\nu}{\vec{q\,}^2} \right)\,,\nonumber\\
P_{\mu \nu}^{E}(q_0,\vec{q\,})
 &= \left(\delta_{\mu\nu}-\frac{q_\mu q_\nu}{q^2} \right)
    -P_{\mu \nu}^{M} (q_0,\vec{q\,})\,.
\label{eq:projectors}
\end{align}


We have fixed our gauge configurations to the minimal Landau gauge by
maximising the gauge fixing functional
\begin{equation}
F[U,g]=\sum_{x,mu}\tr U^g_\mu(x)
 = \sum_{x,\mu}\tr g(x)U_\mu(x)g^\dagger(x+\muhat)\,,
\end{equation}
using the standard overrelaxation algorithm.  The Landau gauge
condition has been imposed with a precision $|\partial_\mu
A_\mu|<10^{-10}$.  We have not investigated the effect of Gribov
copies; this will be left to a future study.

First, in figure~\ref{fig:gluon-vacuum} we show the gluon propagator
and dressing function in the vacuum for our two volumes,
$12^3\times24$ and $16^3\times24$.  Comparing the data for the two
volumes, we see that finite volume effects are modest for these
lattices.  In order to reduce ultraviolet lattice artefacts, we have
applied a weak cylinder cut \cite{Leinweber:1998uu}.  The propagator
exhibits the usual infrared suppression observed in other lattice studies.

In figure~\ref{fig:gluon-compare} we show the two lowest Matsubara
modes for selected spatial momenta as a function of chemical potential
from the $N_\tau=24$ lattices for different spatial volumes.  We find
at most a very mild volume dependence, even for zero spatial momentum,
with some indication that the magnetic propagator is slightly smaller
on the larger volume.  Note that because the available (discrete)
momentum values depend on the spatial volume, the selected momenta
from the $12^3$ and $16^3$ lattices do not match precisely, and the
discrepancy between the propagator values on the two lattices at
nonzero spatial momentum $q_s=|\vec{q}|$ is likely to be at least as
much due to the slightly different values of $q_s$ as to finite volume
effects.

With respect to the
infrared suppressed vacuum propagator shown in
fig.~\ref{fig:gluon-vacuum} we find a mild
enhancement at intermediate $\mu$, i.e.\ in the superfluid, confined
phase, but a suppression in the deconfined phase, i.e.\ for large
$\mu$ for both tensor structures. 

Figure~\ref{fig:gluon-16x8} shows the lowest Matsubara mode for the
high-temperature, $16^3\times8$ lattice, again as a function of
chemical potential and for several different spatial momenta.  Here we
find a considerably more complex picture.  The electric form factor
becomes progressively more suppressed with increasing $\mu$ for all
momenta, while for the magnetic form factor the lowest momentum modes
show an interesting behaviour, with a peak at $\mu a\approx0.5$.
For large spatial momenta this form factor is instead
enhanced at large $\mu$.  

\begin{figure}[thb]
\begin{center}
\includegraphics*[width=\columnwidth]{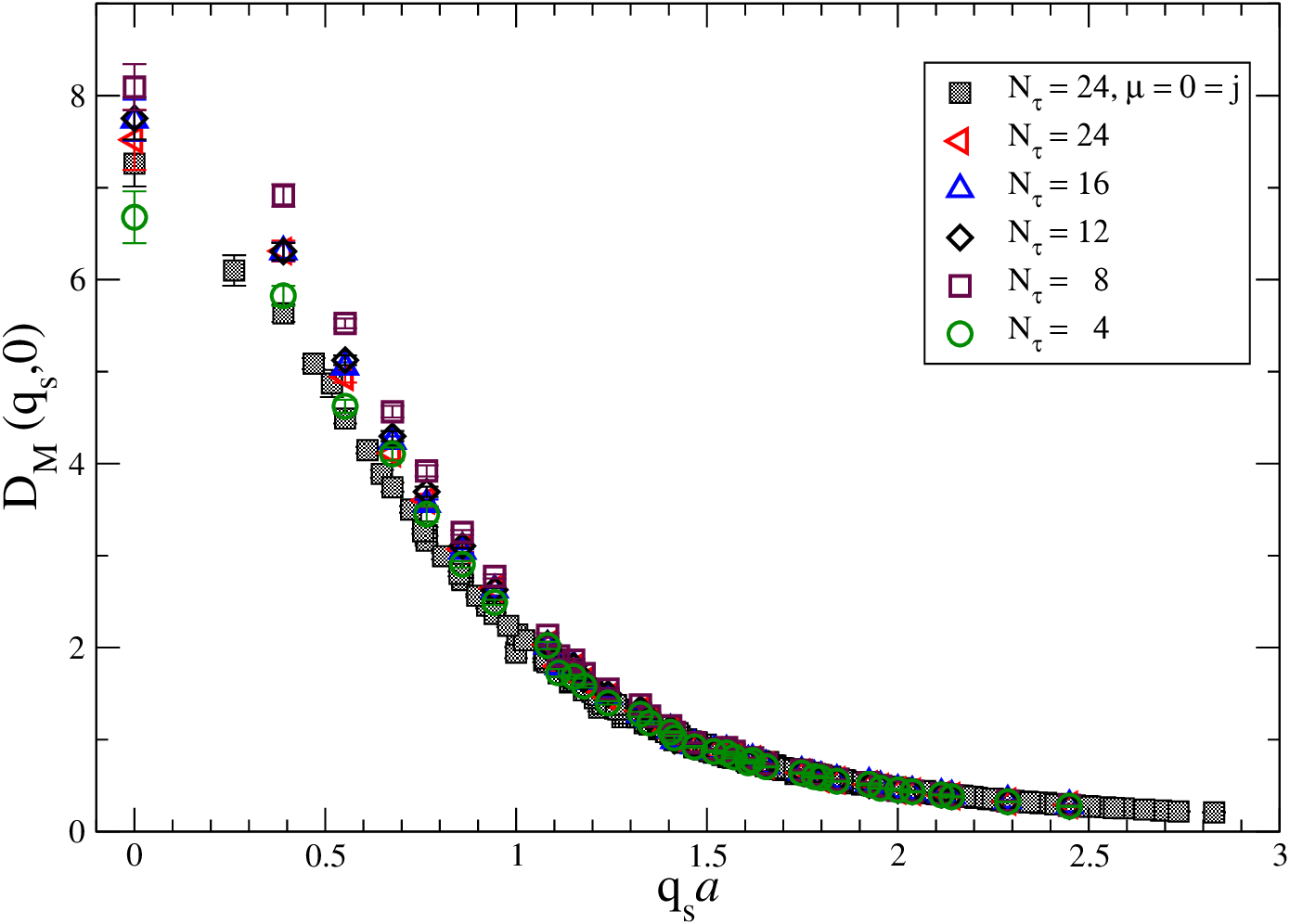}
\includegraphics*[width=\columnwidth]{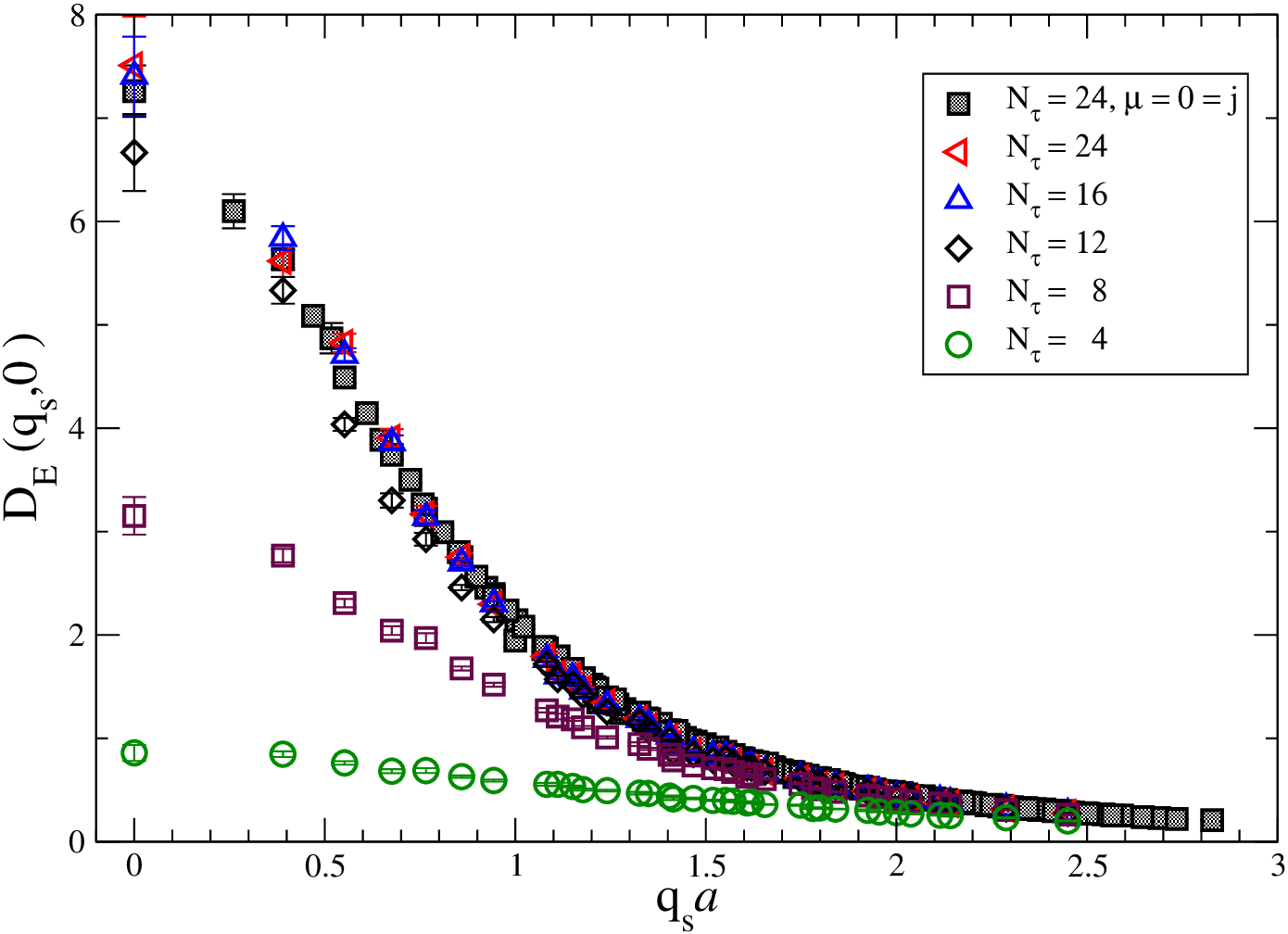}
\caption{Thermal behaviour of the zeroth Matsubara mode of the
  magnetic (top) and electric (bottom) propagators at $\mu a=0.5$ and
  $ja=0.04$ on $16^3\times N_\tau$ lattices.}
\label{fig:zeromodes_vac}
\end{center}
\end{figure}
\begin{figure}[thb]
\begin{center}
\includegraphics*[width=\columnwidth]{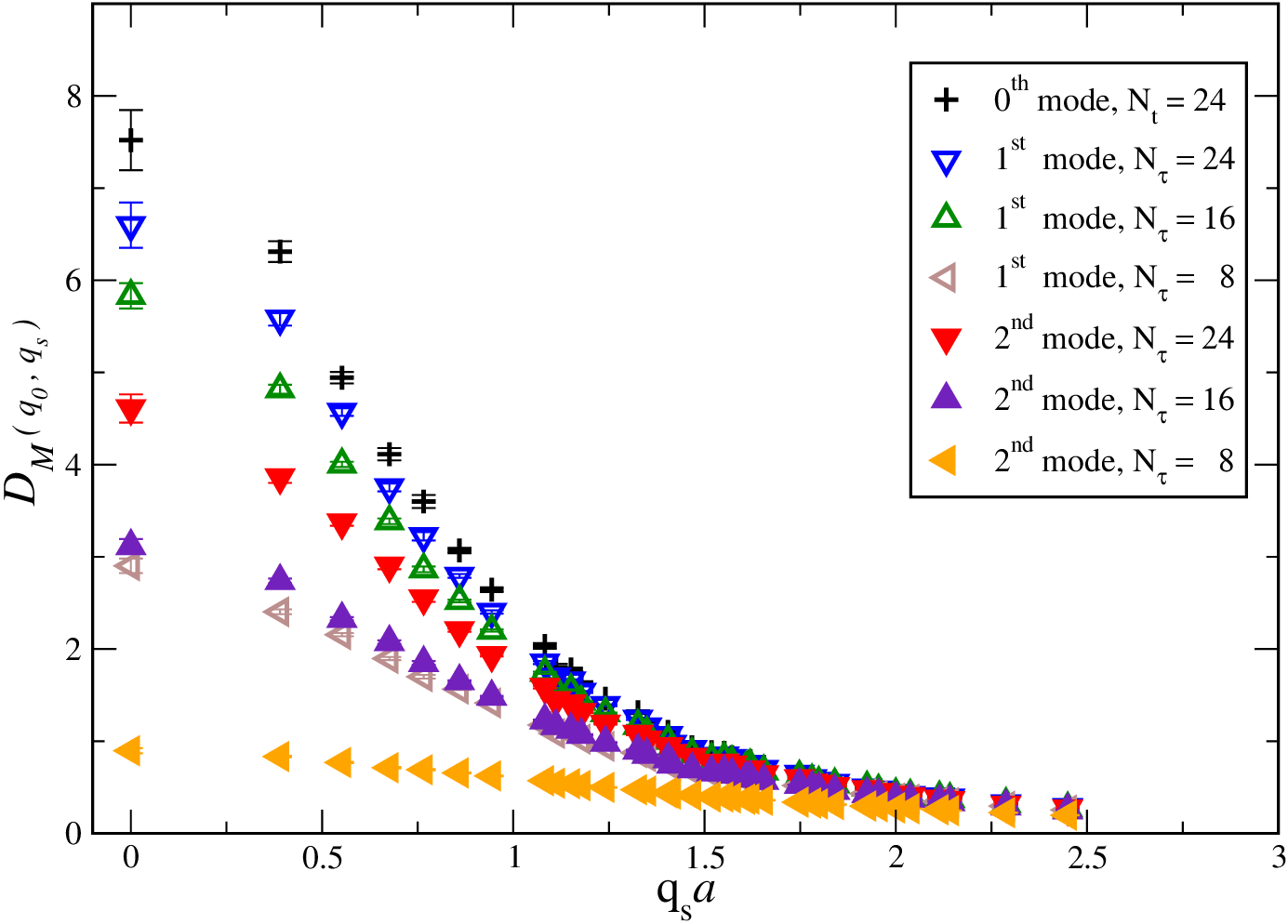}
\includegraphics*[width=\columnwidth]{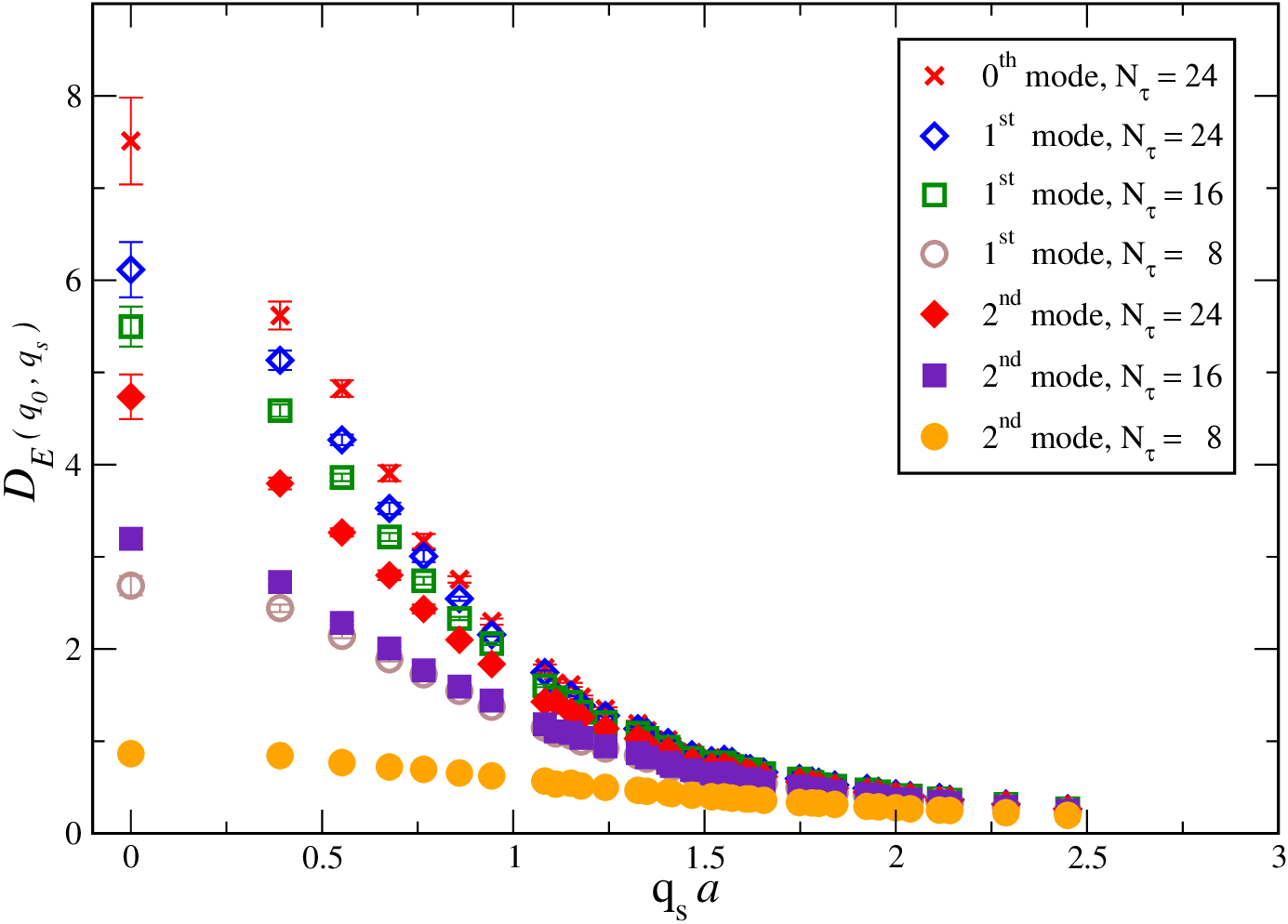}
\caption{Thermal behaviour of higher Matsubara modes of the magnetic
  (top) and electric (bottom) gluon propagators.} 
\label{fig:0th_1st_2nd}
\end{center}
\end{figure}
We now turn to the thermal behaviour of the gluon propagator at fixed
chemical potential.
Fig.\ \ref{fig:zeromodes_vac} shows the zeroth Matsubara modes of the
propagators for $\mu a=0.5$ and $ja=0.04$ on a spatial $16^3$ lattice
as a function of temperature, with higher modes shown in
fig.\ \ref{fig:0th_1st_2nd}. The magnetic component is slightly enhanced by the
effect of chemical potential with respect to the vacuum, but
interestingly it hardly feels thermal effects over a large range of
low to intermediate temperatures. The
electric propagator is suppressed for lower temperatures already. In
pure gauge theory \cite{Cucchieri:2007ta,Fischer:2010fx,Bornyakov:2011jm,Maas:2011se,Aouane:2011fv,Maas:2011ez,Cucchieri:2011di} the zero mode of the electric gluon shows strong
enhancement for temperatures below the deconfinement transition,
whereas the magnetic gluon is screened also below that
transition. However fig.\ \ref{fig:polyakov-allmu} entails that
the chosen value of $\mu$ drives the system close to the phase
transition already at $N_\tau=20$, and hence electric gluon
enhancement may not be observed 
here. The interplay of temperature with chemical potential may also
trigger the observed non-suppression for the magnetic mode. Our
results are however in qualitative agreement with those obtained in a
recent study of QCD with twisted-mass fermions \cite{Aouane:2012bk}. A more
detailed analysis of both medium effects and in particular their
mutual interaction will be presented elsewhere.

\begin{figure}[thb]
\begin{center}
\includegraphics*[width=\columnwidth]{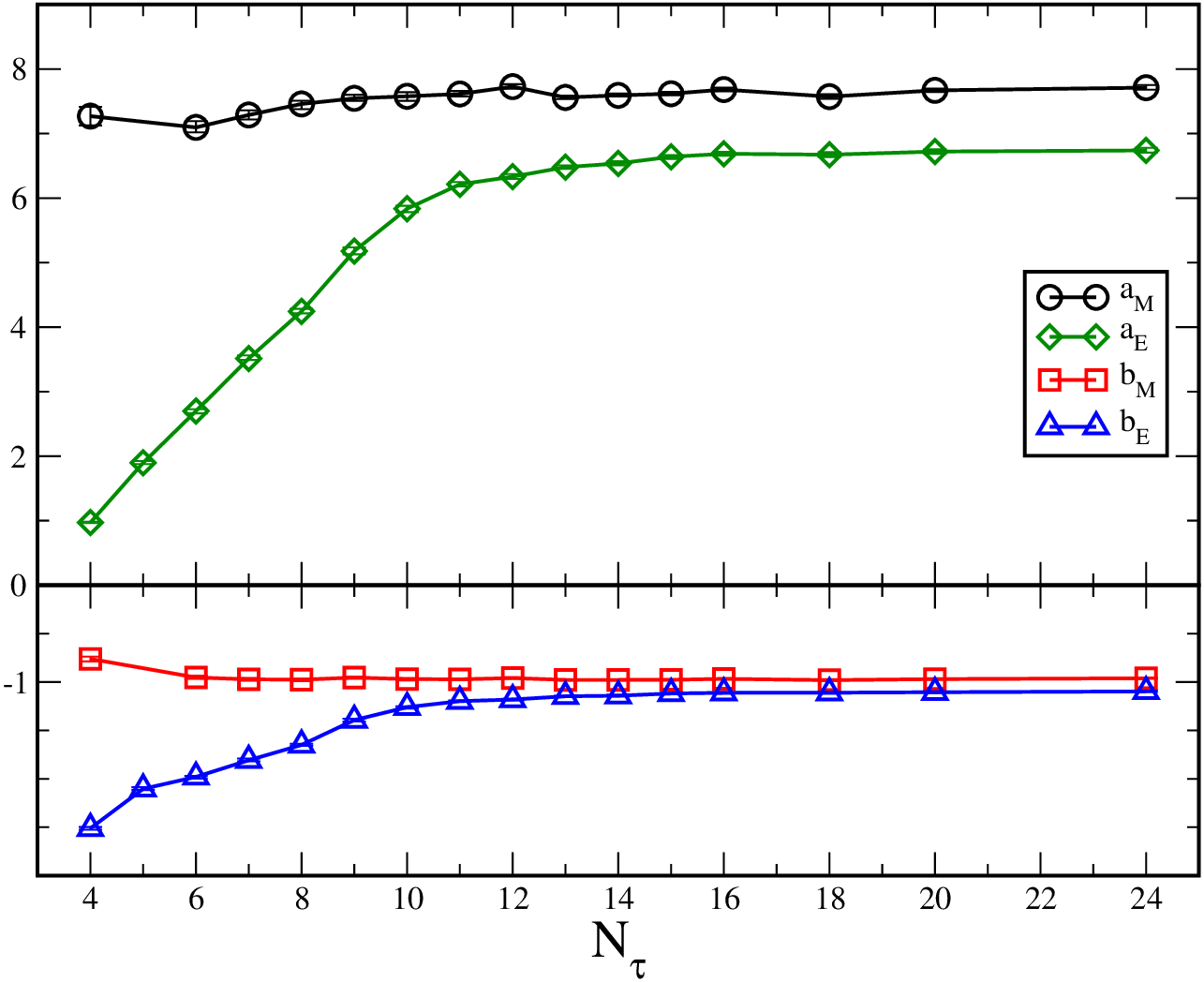}
\caption{$N_\tau$ dependence of fit parameters $a_{M/E}$ and $b_{M/E}$
  for $16^3\times N\tau$ lattices at $\mu a=0.5$ and $ja=0.04$.}
\label{fig:fit_params_T}
\end{center}
\end{figure}

\begin{figure}[thb]
\begin{center}
\includegraphics*[width=\columnwidth]{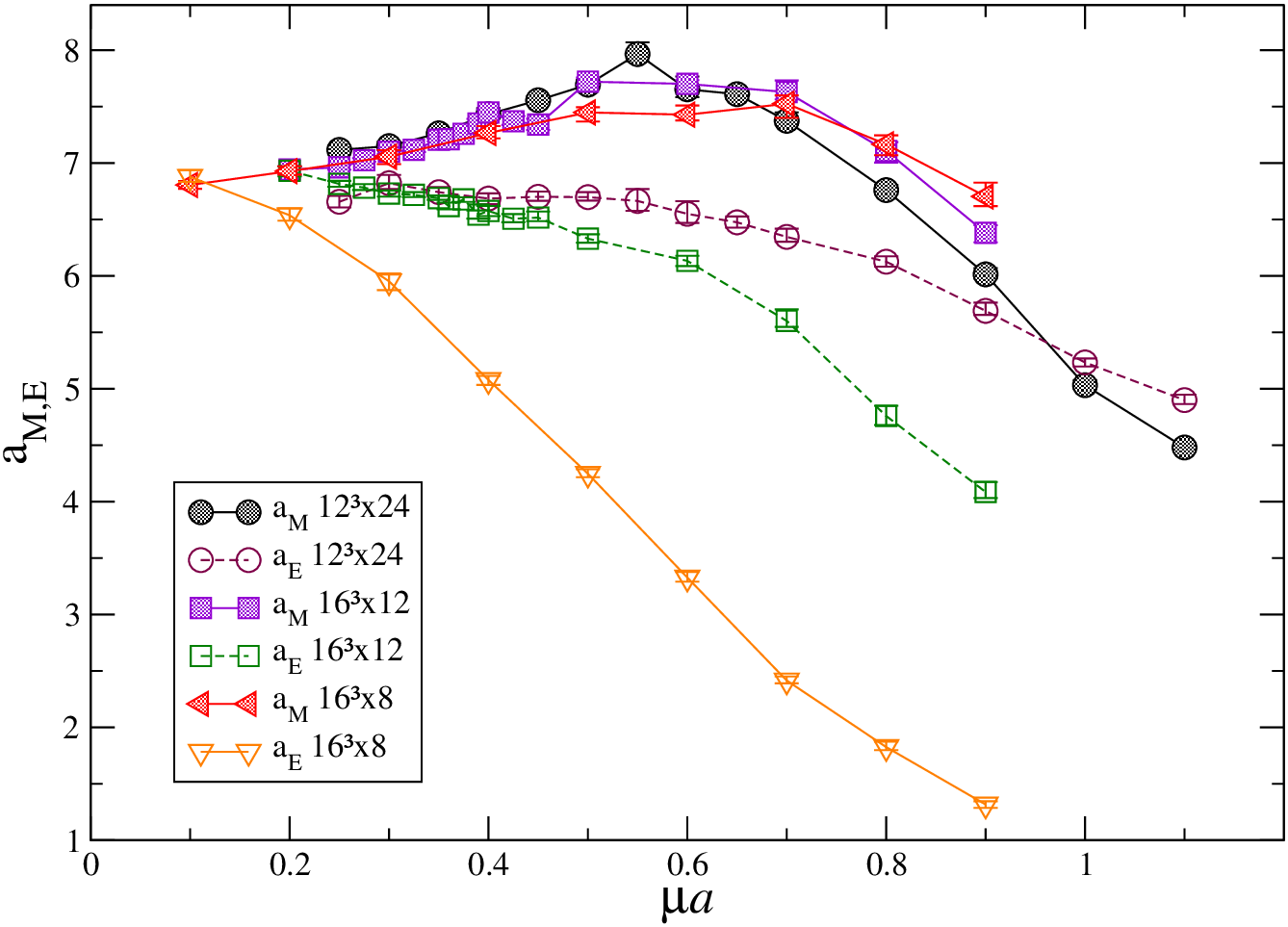}
\includegraphics*[width=\columnwidth]{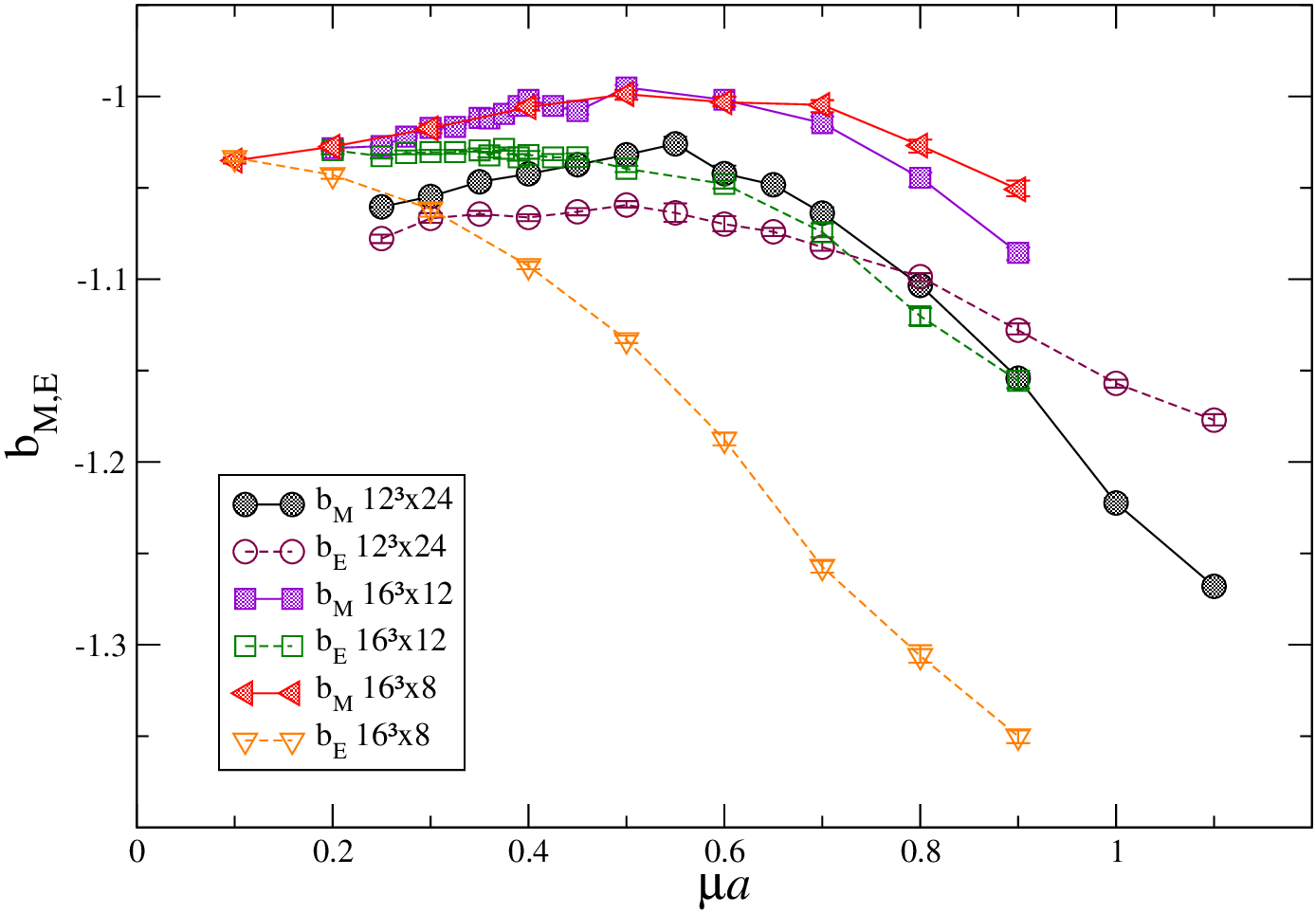}
\caption{Dependence of fit parameters $a_{M/E}$ (top) and $b_{M/E}$ (bottom) on $\mu$ and lattice volume.}
\label{fig:fit_params_mu}
\end{center}
\end{figure}

To further quantify the variation of the gluon propagator with $T$
and $\mu$ we employ a global fit
$D^{\text{fit}}_{M/E}(q^2)$ to all modes, with
$q^2=\vec{q}^2+q_0^2$.  The functional form we use is inspired by
\cite{Fischer:2010fx}. However, for the momenta at hand the
perturbative running of \cite{Fischer:2010fx} can be neglected. In the
vacuum this gives us a three-parameter fit
\begin{equation}
D^{\text{fit}}_{M/E}(q^2) \ = \ \frac{\Lambda^2}{(q^2+\Lambda^2)^2}
\left( q^2+\Lambda^2 a_{M/E} \right)^{-b_{M/E}}\,. 
\label{eq:fit}
\end{equation}
At $\mu\!=\!0\!=\!j$ we find $\Lambda a = 0.999(3)$,
$a_M\!=\!a_E\!=\!6.85(3)$ and $\!b_M\!=\!b_E\!=\!-1.031(2)$ on the
$16^3\times 24$ lattice, with a $\chi^2$ per degree of freedom of
around 8 for the magnetic mode and 5 for the electric mode.  There is
a slight volume dependence, with the $12^3\times24$ lattice yielding
$\Lambda = 0.961(5)$.  The normalisation $\Lambda$ is taken to be
independent of $T$ and $\mu$, but medium effects modify $a_{M/E}$ and
$b_{M/E}$ for magnetic and electric modes individually. The results
for the fit parameters are shown in fig.\ \ref{fig:fit_params_T} as
functions of $N_\tau$ on the $16^3\times N_\tau$ lattices at $\mu a\!=\!0.5$ and
$ja\!=\!0.04$, and in fig.\ \ref{fig:fit_params_mu} as functions of
$\mu$ and lattice volume. For the available data we have found the
dependence on $j$ to be weak.  At $N_\tau=5$ we did not obtain any
satisfactory fit for the magnetic form factor, so these points are
absent from fig.~\ref{fig:fit_params_T}.

\begin{figure}[thb]
\begin{center}
\includegraphics*[width=\columnwidth]{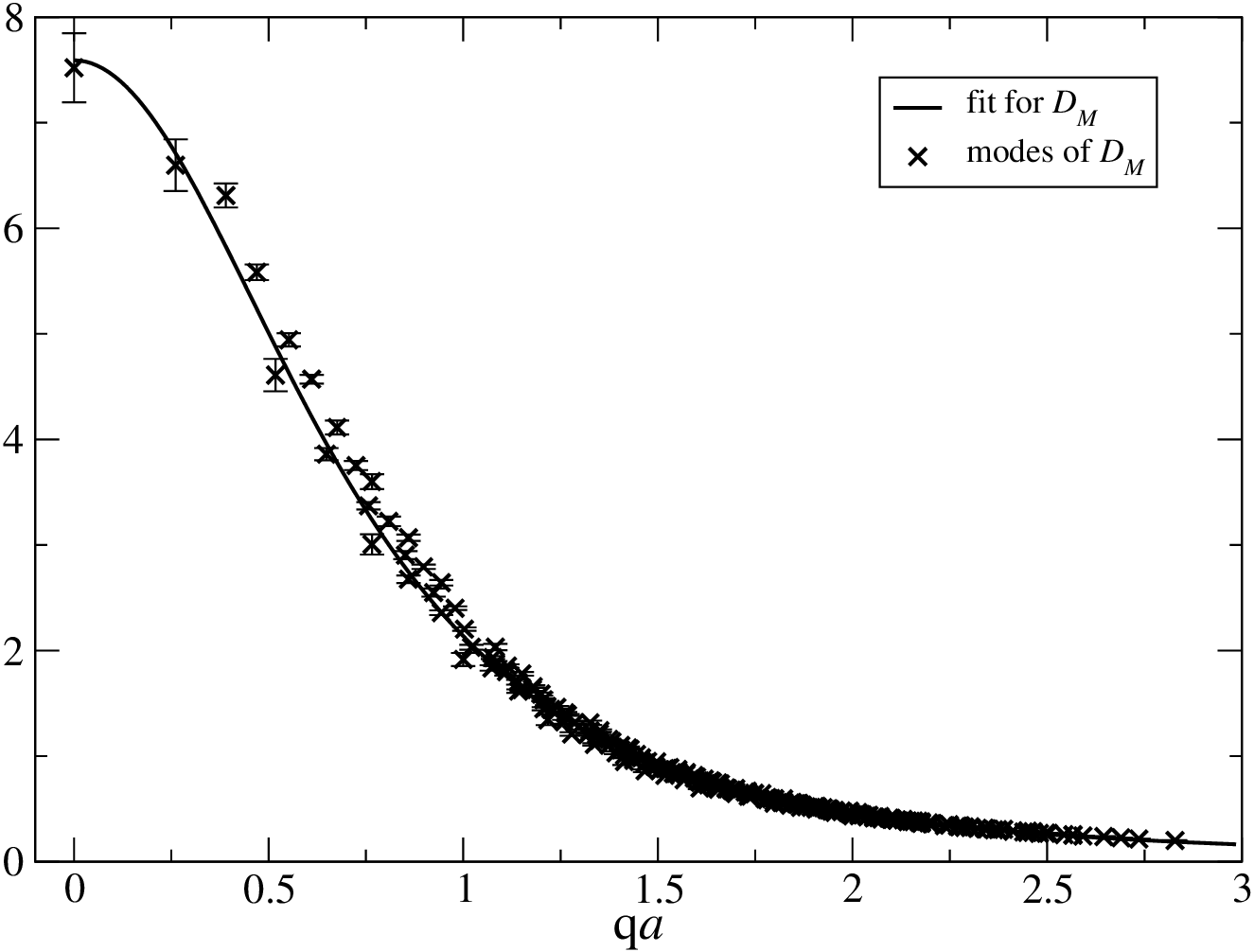}
\includegraphics*[width=\columnwidth]{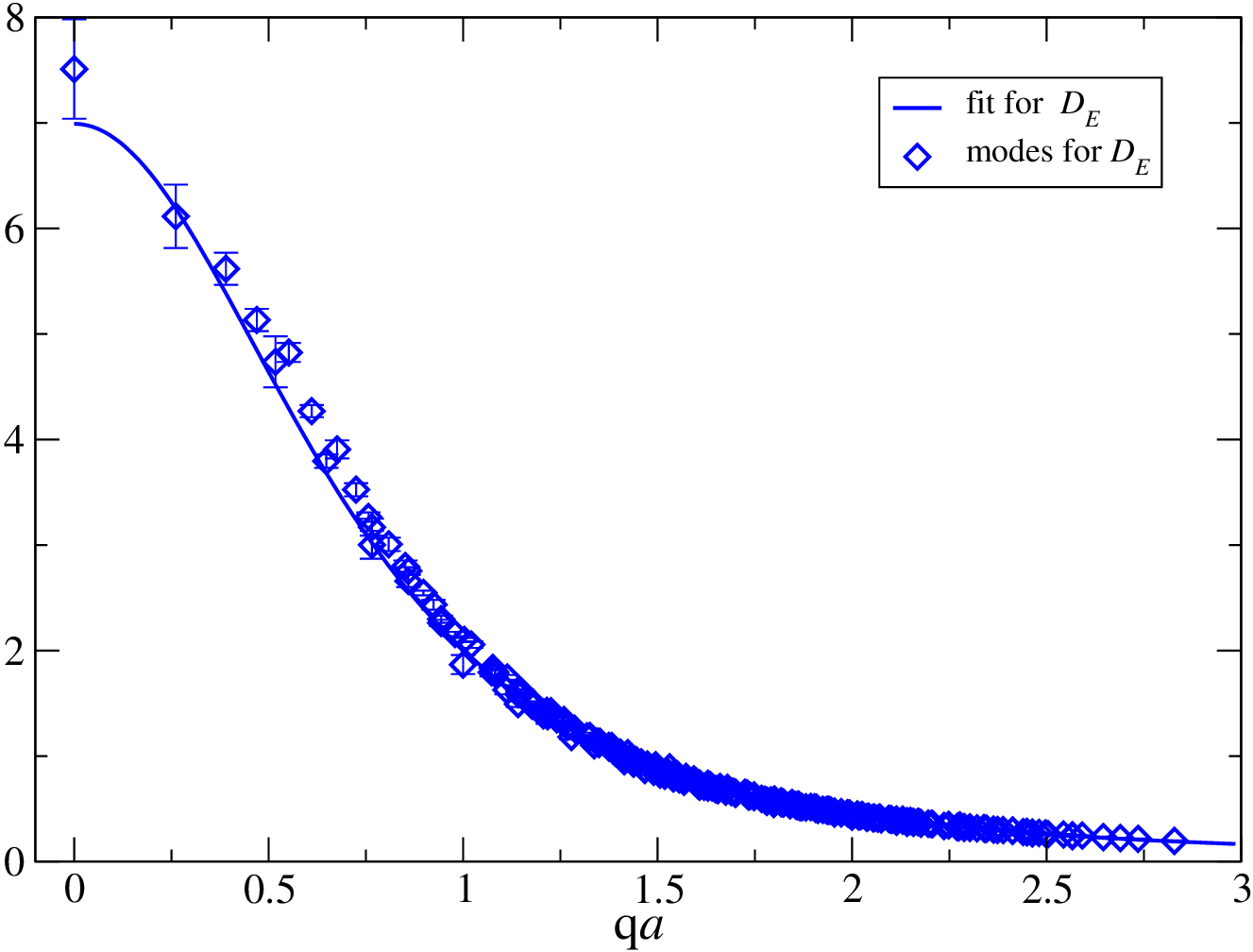}
\caption{Multimodal fits of the form eq.\ (\ref{eq:fit}) for the magnetic (top) and electric (bottom) gluon propagators at $a\mu=0.5$ and $aj=0.04$ on the $16^3\times24$ lattice. Note that the functions are plotted versus four-momentum $q$ on the abscissa.}
\label{fig:fits_mu0500_V16X24}
\end{center}
\end{figure}

To illustrate the quality of these fits,
fig.~\ref{fig:fits_mu0500_V16X24} shows the fits for $\mu a=0.5,
ja=0.04$ on the $16^3\times24$ lattice.  We see that \eqref{eq:fit}
gives a reasonable description for both modes, but the underlying
Ansatz that the gluon propagator is a function of the four-momentum
$q^2$ only, works less well for the magnetic form factor.  The
$\chi^2/N_{df}$ is around 10 for most fits, with the electric form
factor giving in general a somewhat better $\chi^2$.  We are also
planning to employ fit models inspired by hard dense loop perturbation
theory, which depend separately on $q_s$ and $q_0$.

\section{Conclusions}
\label{sec:conclusions}

We have carried out a detailed investigation of several aspects of
2-colour, 2-flavour QCD with $m_\pi/m_\rho\approx0.8$ at nonzero
temperature $T$ and quark chemical potential 
$\mu$.  Our main findings are summarised below.

\begin{enumerate}
\item We have located the superfluid to normal and deconfinement
  transitions in the region $0.35\leq\mu a\leq0.6$
  ($\mu=385-665$ MeV).  The superfluid to normal transition temperature
  $T_s$ is remarkably constant in this region, while the deconfinement
  temperature $T_d$ shows a decrease with $\mu$ which appears to
  continuously connect to the $\mu=0$ transition identified in
  \cphases.  It also appears to extrapolate smoothly to the
  high-$\mu$, low-$T$ transition previously observed \cprev, although
  in the absence on any accurate data for the Polyakov loop at low
  temperature this must be taken merely as indicative.

\item The superfluid to normal transition appears to behave like a
  second order phase transition, while the deconfinement transition
  looks like a smooth crossover, which becomes broader with increasing
  $\mu$.  This would have to be backed up with
  a careful finite volume and critical scaling analysis.

\item The static quark potential at low temperature is at most only
  very weakly screened at large $\mu$, suggesting that the
  dense medium with $\braket{L}\neq0$ is not an ordinary, deconfined
  quark--gluon plasma.

\item The
electric (longitudinal) gluon propagator in Landau gauge becomes
strongly screened with increasing temperature and chemical potential.
The magnetic (transverse) gluon shows little sensitivity to
temperature, and exhibits a mild enhancement at intermediate $\mu$
before coming suppressed at large $\mu$.
\end{enumerate}

\begin{figure}[thb]
\includegraphics*[width=\colw]{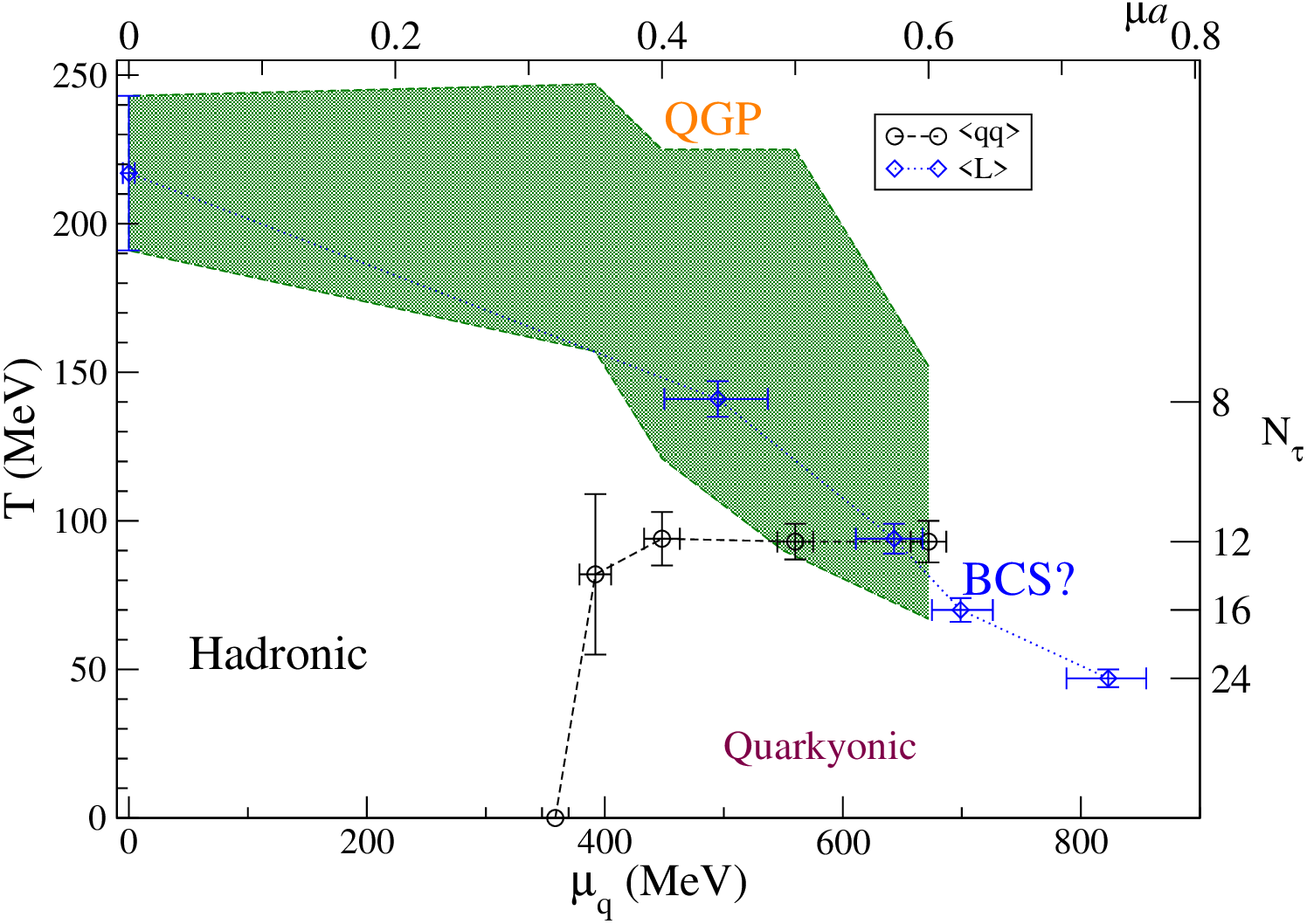}
\caption{Phase diagram of QC$_2$D with $m_\pi/m_\rho=0.8$.  The black
  circles denote the superfluid to normal phase transition; the green
  band the deconfinement crossover.  The blue diamonds are the
  estimates for the deconfinement line from \cphases.}
\label{fig:phasediag}
\end{figure}

The structure of the phase diagram is summarised in
fig.~\ref{fig:phasediag}.  We also include the estimates for the
deconfinement transition given in \cphases.  We see that these are
consistent with our new estimates in this paper.
The indications are that the deconfinement line crosses the superfluid
to normal transition line, giving rise to a region of deconfined,
superfluid matter, but in the absence of precise data at larger $\mu$,
we can not say this with any degree of certainty.
To map out the phase boundaries with greater precision and to clearly
establish the order of the transitions, large-scale simulations on
several spatial volumes will be required.  This goes beyond our
current computational capabilities, but is an interesting topic for
future investigations.

We should also note that these simulations have been carried out with
a rather large quark mass, and it remains to be seen to what extent
these features persist as the quark mass is reduced.

The lack of screening (and possibly even antiscreening) observed in the static
quark potential requires further investigation to establish whether
this is a signal of an exotic state of matter or a result of poor
overlap with relevant states and/or lattice artefacts.
We are planning to compute the static quark potential at higher
temperatures, using both the Wilson loop and Polyakov loop correlator,
to further elucidate this.

The screening of the static magnetic gluon propagator (zero Matsubara
mode) at high $\mu$ and low $T$ is a clear signal of the breakdown of
resummed perturbation theory, which predicts that this mode is
unscreened to all orders.  This is likely to be relevant also for real
QCD, and should be taken into account in any future study using model
gluon propagators to compute for example the superfluid or
superconducting gap.

In an ongoing study we will compare the lattice gluon propagator at
non-vanishing chemical potential and temperature with results from
functional continuum methods, extending studies of thermal propagators
\cite{Fister:2011uw,Fister:2011um,Fister:Diss} to finite density. As
there are no insurmountable problems in ${\rm QC_2D}$, in particular
no sign problem, a direct analysis serves to identify possible
technical limitations in either method, stemming from finite size and
finite volume artefacts in the lattice formulation, or from inevitably
necessary truncations in the continuum description.  We are also in
the process of computing the quark propagator, which will give further
input to these studies.

We are currently extending our study of QC$_2$D to smaller lattice
spacings, which will enable us to perform a controlled extrapolation
to the continuum limit and clarify the possible role of lattice
artefacts at large $\mu$.

\begin{acknowledgement}
This work is carried out as part of the UKQCD collaboration and the
DiRAC Facility jointly funded by STFC, the Large Facilities Capital
Fund of BIS and Swansea University.  We thank the DEISA Consortium
(www.deisa.eu), funded through the EU FP7 project RI-222919, for
support within the DEISA Extreme Computing Initiative.  The simulation
code was adapted with the help of Edinburgh Parallel Computing Centre
funded by a Software Development Grant from EPSRC.  We acknowledge the
use of the USQCD cluster at Fermilab for part of this work.  The work has been
carried out with the support of Science Foundation Ireland grant
11-RFP.1-PHY3193.  DM is supported by U.S. Department of Energy grant
under contract no. DE-FG02-85ER40237.  We thank Pietro Giudice, Simon
Hands and Jan Pawlowski for stimulating discussions and advice.
\end{acknowledgement}

\appendix
\section{Diquark source extrapolation}
\label{appendix}

We have used 3 different functional
forms to extrapolate the diquark condensate to zero diquark source,
\begin{align}
&\text{linear:}&\qquad \qq&=A+Bj\,,\label{eq:linear}\\
&\text{power law:} &\qquad \qq&=Bj^\alpha\,, \label{eq:power}\\
&\text{constant + power:} &\quad \qq&=A+Bj^\alpha\,. \label{eq:powerc}
\end{align}
\begin{table}
  \begin{tabular}{clr|lllc}
 Fit &  $\mu a$ & $N_\tau$ & $A$ & $B$ & $\alpha$ & $\chi^2/N_{df}$ \\ \hline
\eqref{eq:linear}
 & 0.4 &  9 & 0.0036(1) & 0.695(3) & & 26 \\
 & 0.4 & 10 & 0.0056(1) & 0.711(4) & & 102 \\
 & 0.4 & 11 & 0.0096(1) & 0.646(4) & & 9.8 \\
 & 0.5 &  9 & 0.0037(2) & 0.829(4) & & 51 \\
 & 0.5 & 10 & 0.0095(2) & 0.790(6) & & 54 \\
 & 0.5 & 11 & 0.0155(2) & 0.712(6) & & 45 \\
 & 0.6 &  9 & 0.0034(2) & 0.966(4) & & 46 \\
 & 0.6 & 10 & 0.0095(2) & 0.976(6) & & 48 \\
 & 0.6 & 11 & 0.0203(2) & 0.849(6) & & 180 \\ \hline
\eqref{eq:power}
 & 0.4 &  9 & & 0.490(8) & 0.852(5) & 7.4 \\
 & 0.4 & 10 & & 0.438(7) & 0.793(5) & 40 \\
 & 0.4 & 11 & & 0.321(5) & 0.683(5) & 11 \\
 & 0.5 &  9 & & 0.617(10) & 0.875(5) & 28 \\
 & 0.5 & 10 & & 0.439(9) & 0.734(6) & 9.4 \\
 & 0.5 & 11 & & 0.315(6) & 0.612(6) & 1.6 \\
 & 0.6 &  9 & & 0.746(11) & 0.893(5) & 20 \\
 & 0.6 & 10 & & 0.576(9) & 0.767(5) & 19 \\
 & 0.6 & 11 & & 0.357(6) & 0.584(5) & 55 \\ \hline
\eqref{eq:powerc}
 & 0.4 &  9 & -0.003(1) & 0.76(4) & 0.40(3) & 8.7 \\
 & 0.4 & 10 & -0.019(4) & 0.47(4) & 0.24(1) & 3.2 \\
 & 0.4 & 11 & 0.003(3) & 0.85(6) & 0.45(4) & * \\
 & 0.5 &  9 & -0.008(2) & 0.70(4) & 0.42(3) & 29 \\
 & 0.5 & 10 & -0.012(4) & 0.55(5) & 0.31(3) & * \\
 & 0.5 & 11 & -0.005(4) & 0.54(5) & 0.28(2) & 0.88 \\
 & 0.6 &  9 & -0.010(2) & 0.69(3) & 0.48(3) & 0.05 \\
 & 0.6 & 10 & -0.006(4) & 0.68(6) & 0.59(6) & * \\
 & 0.6 & 11 & -0.055(12) & 0.28(4) & 0.269(4) & * \\ \hline
  \end{tabular}
\caption{Parameters for the $j\to0$ extrapolation of $\qq$ in the
  transition region, using a linear \eqref{eq:linear}, power-law
  \eqref{eq:power} and power + constant \eqref{eq:powerc} Ansatz.  All
  four values of $j$ have been used.  Where there is a star in the
  column for $\chi^2/N_{df}$, the central fit value was outside the
  68\% confidence interval, and the quoted value is instead taken to
  be in the middle of the 68\% confidence interval.}
\label{tab:extrapolate}
\end{table}
The results of these extrapolations are summarised in
table~\ref{tab:extrapolate}.  We find that the linear form is clearly
disfavoured; however, the pure power law does not give a good fit
either in most cases, while the constant + power Ansatz is unstable and tends to
give a negative intercept at $j=0$.  Clearly, more work is needed to
obtain good control over the $j\to0$ extrapolation.

%
\bibliography{density,hot,lattice,gluon_prop,paragraph_phases/references}

\begin{thebibliography}{10}

\bibitem{Aarts:2011ax}
G.~Aarts, F.~A. James, E.~Seiler and I.-O. Stamatescu,
\newblock Eur.Phys.J. {\bf C71}, 1756 (2011) [arXiv:1101.3270].

\bibitem{Aarts:2013bla}
G.~Aarts,
\newblock PoS {\bf LATTICE2012}, 017 (2012) [arXiv:1302.3028].

\bibitem{Chandrasekharan:2012fk}
S.~Chandrasekharan,
\newblock Phys.Rev. {\bf D86}, 021701 (2012) [arXiv:1205.0084].

\bibitem{Cristoforetti:2012su}
AuroraScience Collaboration, M.~Cristoforetti, F.~Di~Renzo and L.~Scorzato,
\newblock Phys.Rev. {\bf D86}, 074506 (2012) [arXiv:1205.3996].

\bibitem{Hands:2006ve}
S.~Hands, S.~Kim and J.-I. Skullerud,
\newblock Eur. Phys. J. {\bf C48}, 193 (2006) [hep-lat/0604004].

\bibitem{Hands:2010gd}
S.~Hands, S.~Kim and J.-I. Skullerud,
\newblock Phys. Rev. {\bf D81}, 091502R (2010) [arXiv:1001.1682].

\bibitem{Cotter:2012mb}
S.~Cotter, P.~Giudice, S.~Hands and J.-I. Skullerud,
\newblock Phys.Rev. {\bf D87}, 034507 (2013) [arXiv:1210.4496].

\bibitem{Maas:2012wr}
A.~Maas, L.~von Smekal, B.~Wellegehausen and A.~Wipf,
\newblock Phys.Rev. {\bf D86}, 111901 (2012) [arXiv:1203.5653].

\bibitem{Kogut:2002zg}
J.~B. Kogut and D.~K. Sinclair,
\newblock Phys. Rev. {\bf D66}, 034505 (2002) [hep-lat/0202028].

\bibitem{Hands:2000ei}
S.~Hands {\em et~al.},
\newblock Eur. Phys. J. {\bf C17}, 285 (2000) [hep-lat/0006018].

\bibitem{Hands:2001ee}
S.~Hands, I.~Montvay, L.~Scorzato and J.~Skullerud,
\newblock Eur. Phys. J. {\bf C22}, 451 (2001) [hep-lat/0109029].

\bibitem{Giudice:2004se}
P.~Giudice and A.~Papa,
\newblock Phys.Rev. {\bf D69}, 094509 (2004) [arXiv:hep-lat/0401024].

\bibitem{Cea:2006yd}
P.~Cea, L.~Cosmai, M.~D'Elia and A.~Papa,
\newblock JHEP {\bf 0702}, 066 (2007) [arXiv:hep-lat/0612018].

\bibitem{Cea:2009ba}
P.~Cea, L.~Cosmai, M.~D'Elia, C.~Manneschi and A.~Papa,
\newblock Phys.Rev. {\bf D80}, 034501 (2009) [arXiv:0905.1292].

\bibitem{Kogut:1999iv}
J.~Kogut, M.~Stephanov and D.~Toublan,
\newblock Phys. Lett. {\bf B464}, 183 (1999) [hep-ph/9906346].

\bibitem{Kogut:2000ek}
J.~Kogut, M.~Stephanov, D.~Toublan, J.~Verbaarschot and A.~Zhitnitsky,
\newblock Nucl. Phys. {\bf B582}, 477 (2000) [hep-ph/0001171].

\bibitem{Splittorff:2000mm}
K.~Splittorff, D.~Son and M.~Stephanov,
\newblock Phys. Rev. {\bf D64}, 016003 (2001) [hep-ph/0012274].

\bibitem{Lenaghan:2001sd}
J.~Lenaghan, F.~Sannino and K.~Splittorff,
\newblock Phys.Rev. {\bf D65}, 054002 (2002) [arXiv:hep-ph/0107099].

\bibitem{Splittorff:2001fy}
K.~Splittorff, D.~Toublan and J.~Verbaarschot,
\newblock Nucl.Phys. {\bf B620}, 290 (2002) [arXiv:hep-ph/0108040].

\bibitem{Kogut:2001if}
J.~B. Kogut, D.~Toublan and D.~Sinclair,
\newblock Phys. Lett. {\bf B514}, 77 (2001) [hep-lat/0104010].

\bibitem{Kogut:2002cm}
J.~B. Kogut, D.~Toublan and D.~K. Sinclair,
\newblock Nucl. Phys. {\bf B642}, 181 (2002) [hep-lat/0205019].

\bibitem{Splittorff:2002xn}
K.~Splittorff, D.~Toublan and J.~Verbaarschot,
\newblock Nucl.Phys. {\bf B639}, 524 (2002) [arXiv:hep-ph/0204076].

\bibitem{Dunne:2002vb}
G.~V. Dunne and S.~M. Nishigaki,
\newblock Nucl.Phys. {\bf B654}, 445 (2003) [arXiv:hep-ph/0210219].

\bibitem{Dunne:2003ji}
G.~V. Dunne and S.~M. Nishigaki,
\newblock Nucl.Phys. {\bf B670}, 307 (2003) [arXiv:hep-ph/0306220].

\bibitem{Brauner:2006dv}
T.~Brauner,
\newblock Mod.Phys.Lett. {\bf A21}, 559 (2006) [arXiv:hep-ph/0601010].

\bibitem{Kanazawa:2009ks}
T.~Kanazawa, T.~Wettig and N.~Yamamoto,
\newblock JHEP {\bf 0908}, 003 (2009) [arXiv:0906.3579].

\bibitem{Kanazawa:2009en}
T.~Kanazawa, T.~Wettig and N.~Yamamoto,
\newblock Phys.Rev. {\bf D81}, 081701 (2010) [arXiv:0912.4999].

\bibitem{Kanazawa:2011tt}
T.~Kanazawa, T.~Wettig and N.~Yamamoto,
\newblock JHEP {\bf 1112}, 007 (2011) [arXiv:1110.5858].

\bibitem{Giannini:1991ew}
M.~Giannini, L.~Kondratyuk and M.~Krivoruchenko,
\newblock (1991).

\bibitem{Kondratyuk:1992he}
L.~Kondratyuk and M.~Krivoruchenko,
\newblock Z.Phys. {\bf A344}, 99 (1992).

\bibitem{Rapp:1997zu}
R.~Rapp, T.~Sch{\"a}fer, E.~V. Shuryak and M.~Velkovsky,
\newblock Phys.Rev.Lett. {\bf 81}, 53 (1998) [arXiv:hep-ph/9711396].

\bibitem{Ratti:2004ra}
C.~Ratti and W.~Weise,
\newblock Phys. Rev. {\bf D70}, 054013 (2004) [arXiv:hep-ph/0406159].

\bibitem{Sun:2007fc}
G.-F. Sun, L.~He and P.~Zhuang,
\newblock Phys.Rev. {\bf D75}, 096004 (2007) [arXiv:hep-ph/0703159].

\bibitem{Brauner:2009gu}
T.~Brauner, K.~Fukushima and Y.~Hidaka,
\newblock Phys. Rev. {\bf D80}, 074035 (2009) [arXiv:0907.4905].

\bibitem{Andersen:2010vu}
J.~O. Andersen and T.~Brauner,
\newblock Phys. Rev. {\bf D81}, 096004 (2010) [arXiv:1001.5168].

\bibitem{Harada:2010vy}
M.~Harada, C.~Nonaka and T.~Yamaoka,
\newblock Phys.Rev. {\bf D81}, 096003 (2010) [arXiv:1002.4705].

\bibitem{Zhang:2010kn}
T.~Zhang, T.~Brauner and D.~H. Rischke,
\newblock JHEP {\bf 1006}, 064 (2010) [arXiv:1005.2928].

\bibitem{He:2010nb}
L.~He,
\newblock Phys.Rev. {\bf D82}, 096003 (2010) [arXiv:1007.1920].

\bibitem{Strodthoff:2011tz}
N.~Strodthoff, B.-J. Schaefer and L.~von Smekal,
\newblock Phys.Rev. {\bf D85}, 074007 (2012) [arXiv:1112.5401].

\bibitem{Bloch:1999vk}
J.~C.~R. Bloch, C.~D. Roberts and S.~M. Schmidt,
\newblock Phys. Rev. {\bf C60}, 065208 (1999) [arXiv:nucl-th/9907086].

\bibitem{Nakamura:1984uz}
A.~Nakamura,
\newblock Phys.Lett. {\bf B149}, 391 (1984).

\bibitem{Hands:1999md}
S.~Hands, J.~B. Kogut, M.-P. Lombardo and S.~E. Morrison,
\newblock Nucl. Phys. {\bf B558}, 327 (1999) [hep-lat/9902034].

\bibitem{Kogut:2001na}
J.~B. Kogut, D.~K. Sinclair, S.~J. Hands and S.~E. Morrison,
\newblock Phys. Rev. {\bf D64}, 094505 (2001) [hep-lat/0105026].

\bibitem{Muroya:2002jj}
S.~Muroya, A.~Nakamura and C.~Nonaka,
\newblock Nucl. Phys. Proc. Suppl. {\bf 119}, 544 (2003) [hep-lat/0208006].

\bibitem{Chandrasekharan:2006tz}
S.~Chandrasekharan and F.-J. Jiang,
\newblock Phys.Rev. {\bf D74}, 014506 (2006) [arXiv:hep-lat/0602031].

\bibitem{Alles:2006ea}
B.~All{\'e}s, M.~D'Elia and M.~Lombardo,
\newblock Nucl.Phys. {\bf B752}, 124 (2006) [arXiv:hep-lat/0602022].

\bibitem{Lombardo:2008vc}
M.-P. Lombardo, M.~L. Paciello, S.~Petrarca and B.~Taglienti,
\newblock Eur.Phys.J. {\bf C58}, 69 (2008) [arXiv:0804.4863].

\bibitem{Hands:2011ye}
S.~Hands, P.~Kenny, S.~Kim and J.-I. Skullerud,
\newblock Eur.Phys.J. {\bf A47}, 60 (2011) [arXiv:1101.4961].

\bibitem{Cohen:2003kd}
T.~D.~. Cohen,
\newblock Phys.Rev.Lett. {\bf 91}, 222001 (2003) [arXiv:hep-ph/0307089].

\bibitem{McLerran:2007qj}
L.~McLerran and R.~D. Pisarski,
\newblock Nucl. Phys. {\bf A796}, 83 (2007) [arXiv:0706.2191].

\bibitem{Kojo:2009ha}
T.~Kojo, Y.~Hidaka, L.~McLerran and R.~D. Pisarski,
\newblock Nucl. Phys. {\bf A843}, 37 (2010) [arXiv:0912.3800].

\bibitem{Braun:2007bx}
J.~Braun, H.~Gies and J.~M. Pawlowski,
\newblock Phys.Lett. {\bf B684}, 262 (2010) [arXiv:0708.2413].

\bibitem{Marhauser:2008fz}
F.~Marhauser and J.~M. Pawlowski,
\newblock arXiv:0812.1144.

\bibitem{Braun:2010cy}
J.~Braun, A.~Eichhorn, H.~Gies and J.~M. Pawlowski,
\newblock Eur.Phys.J. {\bf C70}, 689 (2010) [arXiv:1007.2619].

\bibitem{Fister:2013bh}
L.~Fister and J.~M. Pawlowski,
\newblock arXiv:1301.4163.

\bibitem{Skullerud:2008wu}
J.-I. Skullerud,
\newblock Nucl. Phys. {\bf A820}, 175c (2009) [arXiv:0810.3795].

\bibitem{Skullerud:2009hq}
J.-I. Skullerud,
\newblock PoS {\bf QCD-TNT09}, 043 (2009) [arXiv:0912.0844].

\bibitem{Cotter:2012ny}
S.~Cotter {\em et~al.},
\newblock PoS {\bf LATTICE2012}, 091 (2012) [arXiv:1210.6757].

\bibitem{Wilson:1974sk}
K.~G. Wilson,
\newblock Phys.Rev. {\bf D10}, 2445 (1974).

\bibitem{Laine:2006ns}
M.~Laine, O.~Philipsen, P.~Romatschke and M.~Tassler,
\newblock JHEP {\bf 0703}, 054 (2007) [arXiv:hep-ph/0611300].

\bibitem{Brambilla:2008cx}
N.~Brambilla, J.~Ghiglieri, A.~Vairo and P.~Petreczky,
\newblock Phys.Rev. {\bf D78}, 014017 (2008) [arXiv:0804.0993].

\bibitem{Beraudo:2007ky}
A.~Beraudo, J.-P. Blaizot and C.~Ratti,
\newblock Nucl.Phys. {\bf A806}, 312 (2008) [arXiv:0712.4394].

\bibitem{Beraudo:2010tw}
A.~Beraudo, J.~Blaizot, P.~Faccioli and G.~Garberoglio,
\newblock Nucl.Phys. {\bf A846}, 104 (2010) [arXiv:1005.1245].

\bibitem{Bazavov:2012bq}
A.~Bazavov and P.~Petreczky,
\newblock Nucl.Phys.A904-905 {\bf 2013}, 599c (2013) [arXiv:1210.6314].

\bibitem{Kawanai:2011jt}
T.~Kawanai and S.~Sasaki,
\newblock Phys.Rev. {\bf D85}, 091503 (2012) [arXiv:1110.0888].

\bibitem{Allton:2012ki}
C.~Allton, W.~Evans and J.-I. Skullerud,
\newblock PoS {\bf LATTICE2012}, 082 (2012) [arXiv:1306.3140];
%
\newblock arXiv:1303.5331.

\bibitem{Hands:2012yy}
S.~Hands, S.~Kim and J.-I. Skullerud,
\newblock Phys.Lett. {\bf B711}, 199 (2012) [arXiv:1202.4353].

\bibitem{Bazavov:2013zha}
A.~Bazavov and P.~Petreczky,
\newblock arXiv:1303.5500.

\bibitem{Bali:2005fu}
SESAM Collaboration, G.~S. Bali, H.~Neff, T.~D{\"u}ssel, T.~Lippert and
  K.~Schilling,
\newblock Phys.Rev. {\bf D71}, 114513 (2005) [arXiv:hep-lat/0505012].

\bibitem{Leinweber:1998uu}
UKQCD, D.~B. Leinweber, J.~I. Skullerud, A.~G. Williams and C.~Parrinello,
\newblock Phys. Rev. {\bf D60}, 094507 (1999) [hep-lat/9811027].

\bibitem{Cucchieri:2007ta}
A.~Cucchieri, A.~Maas and T.~Mendes,
\newblock Phys.Rev. {\bf D75}, 076003 (2007) [arXiv:hep-lat/0702022].

\bibitem{Fischer:2010fx}
C.~S. Fischer, A.~Maas and J.~A. Muller,
\newblock Eur. Phys. J. {\bf C68}, 165 (2010) [arXiv:1003.1960].

\bibitem{Bornyakov:2011jm}
V.~Bornyakov and V.~Mitrjushkin,
\newblock Int.J.Mod.Phys. {\bf A27}, 1250050 (2012) [arXiv:1103.0442].

\bibitem{Maas:2011se}
A.~Maas,
\newblock Phys. Rep. in press  (2013) [1106.3942].

\bibitem{Aouane:2011fv}
R.~Aouane {\em et~al.},
\newblock Phys.Rev. {\bf D85}, 034501 (2012) [arXiv:1108.1735].

\bibitem{Maas:2011ez}
A.~Maas, J.~M. Pawlowski, L.~von Smekal and D.~Spielmann,
\newblock Phys.Rev. {\bf D85}, 034037 (2012) [arXiv:1110.6340].

\bibitem{Cucchieri:2011di}
A.~Cucchieri and T.~Mendes,
\newblock PoS {\bf FACESQCD}, 007 (2010) [arXiv:1105.0176].

\bibitem{Aouane:2012bk}
R.~Aouane, F.~Burger, E.-M. Ilgenfritz, M.~Muller-Preussker and A.~Sternbeck,
\newblock Phys. Rev. D 87, {\bf 114502} (2013) [arXiv:1212.1102].

\bibitem{Fister:2011uw}
L.~Fister and J.~M. Pawlowski,
\newblock arXiv:1112.5440.

\bibitem{Fister:2011um}
L.~Fister and J.~M. Pawlowski,
\newblock arXiv:1112.5429.

\bibitem{Fister:Diss}
L.~Fister,
\newblock PhD thesis, Heidelberg University  (2012).

\end{thebibliography}

\end{document}